\documentclass[12pt,preprint]{aastex}

\newcommand{\et}{et al.} 
\newcommand{\kms}{km s$^{-1}$}
\newcommand{\ha}{H$\alpha$}
\newcommand{\solar}{\ifmmode_{\sun}\else$_{\sun}$\fi}
\newcommand{\msun}{M$_{\sun}$}
\newcommand{\ergsec}{ergs s$^{-1}$} 
\newcommand{\HII}{H$\,${\sc ii}}
\newcommand{\HI}{H$\,${\sc i}} 
\newcommand{\coldens}{cm$^{-2}$}
\newcommand{\acm}{cm$^{-2}$}
\newcommand{\cd}{C$+$D}
\newcommand{\cdu}{C$+$D$_{UN}$}
\newcommand{\cdn}{C$+$D$_{NA}$}

\begin{document}

\slugcomment{To appear in The Astronomical Journal}


\title{DDO 43: A Prototypical Dwarf Irregular Galaxy?}

\author{Caroline E.\ Simpson}
\affil{Department of Physics, Florida International University,
  University Park, Miami, Florida 33199 USA} 
\email{simpsonc@galaxy.fiu.edu}

\author{Deidre A.\ Hunter}
\affil{Lowell Observatory, 1400 West Mars Hill Road, Flagstaff, Arizona
   86001 USA}
\email{dah@lowell.edu}

\and

\author{Tyler E.\ Nordgren}
\affil{Department of Physics, University of Redlands, 1200 East Colton
  Avenue, Redlands, CA 92373 USA}
\email{tyler\_nordgren@redlands.edu}

\begin{abstract}

We present sensitive and high resolution 21 cm observations of the dwarf
irregular (dIm) galaxy DDO 43, in conjunction with optical broad and narrow
band images in U, B, V, and \ha. The observations are used to examine
the relationship between its \HI\ morphology and kinematics to past and
present star formation. Optically, it is a small (R$_{25}$ = 990 pc),
faint (M$_B$ of $-$14.0) Im with a slightly boxy shape.  In \HI, DDO 43
has an extended (R$_{HI}$/R$_H = 2.8$) gas envelope. There is a high
density ridge associated with the optical body of the galaxy, containing
several higher density knots and lower density holes. The largest hole
is $\sim 850\times 530$ pc. No expansion is detected, so it must be
relatively old.  The largest and potentially oldest (7--70 Myr) of the
six identified star clusters is located at the western edge of the
hole. Four of the other clusters are located near high density peaks.
There are several \HII\ regions, most (but not all) of which are
associated with peaks in the \HI\ surface density. The overall star
formation rate is average for its type.

In many ways, DDO 43 is a very typical dwarf irregular galaxy. Its \HI\
morphology is consistent with a history of episodes of localized star
formation that create holes and shells in the ISM, some of which can
overlap. These features are located within the area of solid-body
rotation in the galaxy; the lack of shear in these small systems allows
such structures to persist for long periods of time.

\end{abstract}


\section{Introduction}

DDO 43 is part of a large survey of Im galaxies being conducted by
Hunter and Elmegreen (2004). The survey includes 140 Im, Sm, and Blue
Compact Dwarf (BCD) galaxies, thus we can place DDO 43 in the context of
a large sample.  One of the purposes of the survey is to examine how star formation
occurs and is regulated in low mass galaxies. The galaxies were chosen to be relatively nearby, and
the sample was biased to those systems containing gas.  All galaxies
have been observed in \ha\ and UBV, and a sub-sample in JHK and \HI\
(neutral hydrogen).
 For those observed in \HI,
we are examining the relationship between the stellar and gaseous
components of galaxies in the sample. Here we present the results for
the Im galaxy DDO 43.

Optically, DDO 43 shows no hint of being different from other tiny dwarf
galaxies. It looks rather rectangular in V, as many Im galaxies do, and
it contains numerous \HII\ regions.  At a distance of 5.5 Mpc
(determined from V$_{GSR}$ given by de Vaucouleurs \et\ 1991 [$=$RC3]
and a Hubble constant of 65 \kms\ Mpc$^{-1}$), DDO 43 has an integrated
M$_B$ of $-$14.0.  This places DDO 43 towards the faint end of the
distribution of Im galaxies, but still well within normal bounds (Figure
3 of Hunter 1997).  The oxygen abundance of DDO 43 was measured by
Hunter \& Hoffman (1999) to be $\log O/H + 12 = 8.3\pm0.09$. This value
is typical of Im galaxies (see Figure 5 of Hunter \& Hoffman).  DDO 43
is relatively isolated. Its nearest neighbor is another Im galaxy, DDO 46,
that is 270 kpc away.  The nearest large galaxy is the Sc spiral NGC
2537A 1.6 Mpc away, including the radial difference.

Despite its optical normality, there were two possible characteristics
of DDO 43 that led us to examine it in more detail.  Our discovery of a
large (galaxy-sized) hole in the \HI\ of the dwarf Im galaxy DDO 88
(Simpson, Hunter, \& Knezek 2005) prompted us to further explore DDO
43. Both DDO 88 and DDO 43 were originally observed using the Very Large
Array (VLA)\footnote{The VLA is a facility of the National Radio
Astronomy Observatory, operated by Associated Universities, Inc., under
cooperative agreement with the National Science Foundation.} while it
was in its C configuration. Although this produced data with fairly good
resolution, these data suffered from low signal-to-noise, and only the
highest density \HI\ peaks in each galaxy were detected. For both DDO 88
and DDO 43, these peaks appeared to form an incomplete shell around the
optical body of the galaxy (Simpson \& Gottesman 2000). Intrigued by
this, we obtained more sensitive \HI\ data for both galaxies with the
VLA in both C and D-configuration. These deeper images of DDO 88
revealed an extensive \HI\ envelope, and a high-density ring of \HI\
surrounding the bulk of the optical body (Simpson \et\ 2005). The inner
part of the ring is \HI-deficient, and the size of the hole is
comparable to the optical galaxy. There is a faint, very red cluster
located approximately in the center of the hole that may be the remnant
of the star-formation event that produced the \HI\ hole.

Previous work suggests that feedback between energy deposited in the
interstellar medium (ISM) by massive star formation and the ISM itself
may regulate star formation in low mass dwarf galaxies. They do not possess
the gravitational potentials necessary to support spiral density waves,
which regulate star formation in larger galaxies, so some other
mechanism must be operating. \HI\ observations of dIrrs have
revealed highly structured \HI\ morphologies, with holes, shells, and
filaments apparent (e.g.\ Ott \et\ 2001; Stewart \et\ 2000; Walter \&
Brinks 1999; Martin 1998; Meurer \et\ 1998; Young \& Lo 1996; Meurer
\et\ 1992; Puche \et\ 1992; but see Rhode \et\ 1999). Presumably, these
features are both the result and cause of episodes of star formation:
energy and stellar winds from massive stars sweeps up the surrounding
\HI\, producing cavities in the gas (e.g.\ Scalo \& Chappell
1999). Denser regions of gas pile up along the boundaries, creating \HI\
shells and filaments. Where the gas becomes dense enough, further star
formation can occur.

The detection of an old, red cluster in the center of the large \HI\
hole in DDO 88, along with the location of younger clusters in the
higher density shell around the hole, is in agreement with this star
formation scenario. These interesting results from our study of DDO 88
led us to examine DDO 43 to see if it also contained a large \HI\ hole,
as hinted at by the earlier VLA observations. DDO 43 is relatively
similar to DDO 88 in many physical respects; has star formation had the
same sort of effect on it as in DDO 88? 

DDO 43 also interested us because it is listed by Hunter, Hunsberger, \&
Roye (2000) as a candidate old tidal dwarf galaxy.  People have recently
recognized that dwarf irregular (Im) galaxies can form in the tidal
tails of merging spiral galaxies (Mirabel, Dottori, \& Lutz 1992;
Malphrus \et\ 1997; Duc \& Mirabel 1999).  Some dwarf systems are
forming this way now (Hunsberger, Charlton, \& Zaritsky 1996) while
others may have formed in interactions long ago that are no longer
obvious.  A distinctive characteristic of tidal dwarfs is that they
should have less dark matter than traditional dwarfs (Barnes \&
Hernquist 1992). Hunter \et\ exploited this property to identify
candidate old tidal dwarf systems among nearby Im galaxies.  DDO 43
potentially has a maximum rotation speed that is too low for its
luminosity (Simpson \& Gottesman 2000; Figure 1 of Hunter \et\ 2000). If
borne out, this could indicate a deficit of dark matter and make DDO 43
a strong candidate for a tidal dwarf.

\section{Data Acquisition and Reductions}

\subsection{Optical}

We used the Hall 1.1 m telescope at Lowell Observatory to obtain B and V
images of DDO 43 1999 February and U images 1999 November.  We used a
SITe 2048$\times$2048 CCD binned $2\times2$.  The exposure times were
$6\times900$ s for V, $5\times1800$ s for B, and $3\times1800$ s for
U. The pixel scale was 1.13\arcsec, and the seeing in the final averaged
images was 2.9\arcsec\ in V, 2.3\arcsec\ in B, and 2.6\arcsec\ in U.
The electronic pedestal was subtracted using the overscan strip, and the
images were flat-fielded using sky flats.  Landolt (1992) standard stars
were used to calibrate the photometry.  The images in each filter were
averaged with an algorithm to reject cosmic rays while still preserving
the photometric integrity of the image.

Before performing surface photometry, we edited foreground stars and
background galaxies from the final UBV images. The V-band image was done
by hand, and then the cursor log file produced from that was used to
remove objects from the other two filters. In that way, stars were
removed in the same way for each filter.  We then made a two-dimensional
fit to the background and subtracted it from the image to produce a
sky-subtracted final image for analysis.

We obtained \ha\ images of DDO 43 in 1995 February with the Perkins 1.8 m
telescope at Lowell Observatory. We used a TI $800\times800$ CCD
provided to Lowell Observatory by the National Science Foundation and
the Ohio State University Fabry-Perot as a 3:1 focal reducer.  The \ha\
filter has a central wavelength of 6566 \AA\ and a FWHM of 32 \AA.  The
contribution from stellar continuum was determined using an off-band
filter centered at 6440 \AA\ with a FWHM of 95 \AA.  The off-band image
was shifted, scaled, and subtracted from the \ha\ image to remove the
stellar continuum and leave only \ha\ nebular emission.  The \ha\
observations consisted of three 3000 s images, preceeded by a 900 s
off-band observation.  The pixel scale was 0.49\arcsec, and the stellar
FWHM on the final image is 1.6\arcsec.  The electronic pedestal was
subtracted using the overscan strip, and the images were flat-fielded
using observations of a white spot hanging on the inside of the dome.

The \ha\ flux was calibrated using observations of the NGC 2363 nebula
(Kennicutt, Balick, \& Heckman 1980).  The \ha\ flux has been corrected
for the shift of the bandpass with temperature and for contamination by
the [NII]$\lambda$6584 emission line in the bandpass (a 1\% correction).

\subsection{\HI\ Data} \label{HIdata}

21 cm spectral line observations of DDO 43 were made using both the D
and C configurations of the VLA. Time on source for both observations
was approximately 4.5 hours. The D configuration, which is the VLA's
most compact, provides short spacings, resulting in high sensitivity but
at the cost of spatial resolution: the synthesized beam is on the order
of 1\arcmin. The C configuration, with longer baselines, provides higher
resolution (nominally 12.5\arcsec) but lower sensitivity. Observations
for both configurations were made using a 128 channel spectrometer in
2IF mode with on-line Hanning smoothing. This results in a channel
separation of 12.2 kHz (2.6 \kms) for a total observed bandwidth of 1.56
MHz. For spectral line data that are taken using on-line Hanning
smoothing, the velocity resolution is equal to the channel separation,
and so is also 2.6 \kms.

Observations of the nearby continuum source 0713+438 (J2000) were made
for phase calibration of both data sets; observations of the sources
0137+331 (J2000) and 1331+305 (J2000) were made for calibration of the
flux and bandpass of the C and D-configuration data, respectively. Calibration
of both data sets was performed using the standard routines in
AIPS. Observational information is provided in Table~\ref{tab:HI}.

\placetable{tab:HI}


The D-configuration source data were affected by solar contamination on the
short baselines. The corrupted data were flagged, removing approximately
8\% of the total number of visibilities. There were three strong
continuum sources in the field; there was no sign of continuum emission
from DDO 43. The continuum was satisfactorily
subtracted in the {\it uv} plane by using line-free channels to make a
linear fit to the continuum emission. The resulting continuum-free data
was then imaged using the AIPS task {\scshape imagr}. To ensure the best
sensitivity to low levels of emission, natural weighting was used. The
data were {\scshape clean}ed during imaging to reduce the sidelobes
produced by non-gaussian features in the ``dirty'' beam. No clean boxes
were specified, so the default of a box ten pixels smaller than the
image size was used. No zero-spacing flux was assumed. The specified
flux cutoff of 0.83 mJy/B ($\sim 1\sigma$ in a single channel) was 
reached in less than 3000 iterations in each channel, which is less than
the 10,000 iteration cutoff set in the task.

The resulting data cube (one channel for each observed frequency) has a
beam size of $54.8\arcsec\times 53.7\arcsec$\ and a single channel
r.m.s.\ of 0.81 mJy/B. Line emission appears from approximately 300--420
\kms. The data cube was integrated in velocity with the task {\sc
momnt} to produce moment maps of the integrated flux density, the
flux-weighted velocity field, and the velocity dispersions. The
integration task first smooths the data spatially and then in velocity
space to create a lower-noise mask. A cutoff of approximately 1.5 times
the single-channel r.m.s.\ was applied to the mask, and pixels in
the original, unsmoothed cube corresponding to the mask pixels higher
than the cutoff flux were integrated.


The C-configuration data were very well behaved resulting in minimal data
editing.  No solar contamination was present, but like the D configuration,
there were still the three strong continuum sources.  This continuum
emission was subtracted as for the D-configuration data. The imaging was done
using uniform weighting ({\sc robust} = 1) however, to produce high
resolution images with only a slight degredation in sensitivity. This
resulted in a beam size of $14.0\arcsec\times 11.0\arcsec$\ and a
single-channel r.m.s.\ of 1.05 mJy/B. As for the D-configuration data,
the data were {\sc clean}ed during imaging. In this case, a test showed
that the {\sc clean} component flux leveled off after the first 500
iterations, and so an iteration cutoff of 2500 was used in conjunction
with a flux cutoff of 0.6 mJy/B ($\sim 0.5\sigma$). The same defaults of
no zero-spacing flux and no specified {\sc clean} boxes were used.
After imaging, the cube was integrated in velocity using a flux cutoff
of $1.5\sigma$ to produce moment maps as was done for the D-configuration data.


Once the C and D-configuration data were calibrated, they were combined in the
{\it uv} plane to produce a data set with both good resolution and
sensitivity. This combined \cd\ dataset was imaged using both natural
(\cdn) and uniform (\cdu; {\sc robust} = 1) weighting. The
naturally-weighted data cube has a beam size of $24.5\arcsec\times
17.1\arcsec$ and a single-channel r.m.s.\ of 1.05 mJy/B. The
uniformly-weighted cube has a beam size of $20.30\arcsec\times
14.00\arcsec$ and a single-channel r.m.s.\ of 1.07 mJy/B (see Table
\ref{tab:HI}). As for the C configuration data, an iteration cutoff of
2500 was determined to be adequate, and was combined with a flux cutoff
of 0.6 mJy/B ($\sim .5\sigma$) for the uniform weighted data. For the
naturally weighted data, the interation limit was set to 3500. No clean
boxes or zero-spacing flux were specified.

\begin{figure}
\epsscale{0.8}
\plotone{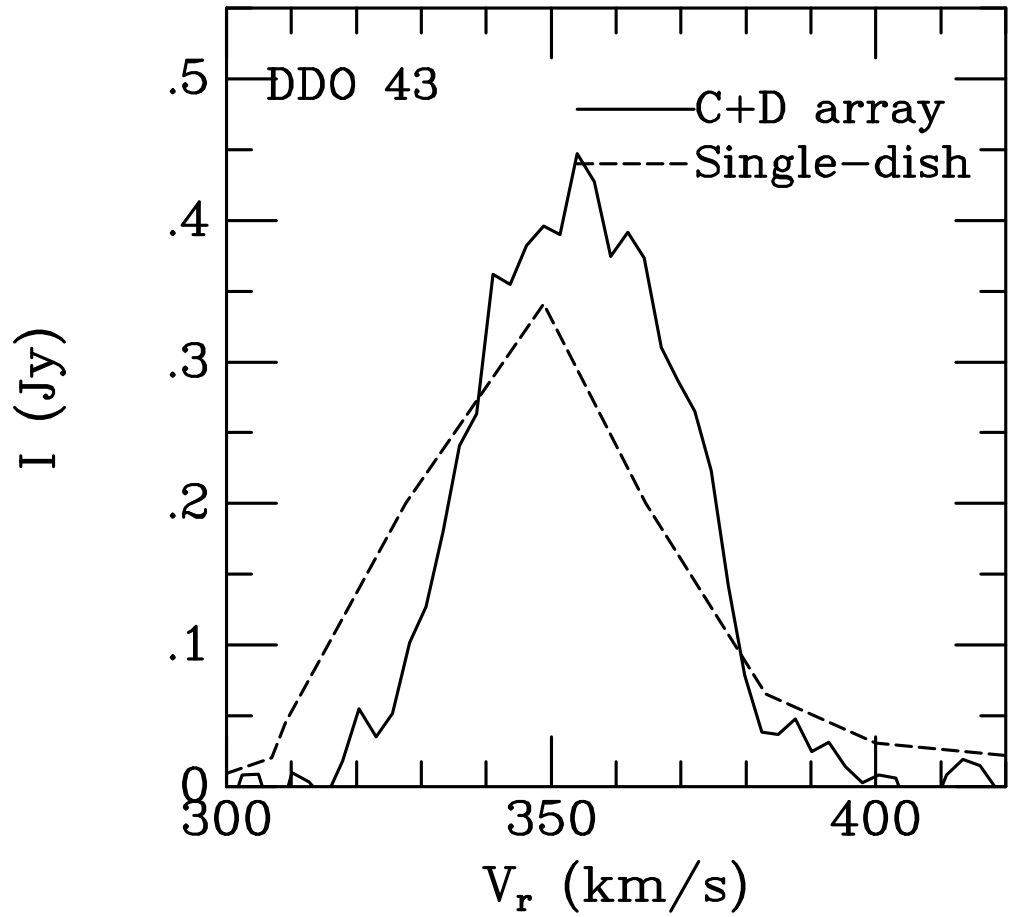}
\caption{Velocity profiles from our
C$+$D-configuration 
data (natural weighting) and from single-dish data (Huchtmeier \&
Richter 1986). 
\label{fig:single}}
\end{figure}

In Figure~\ref{fig:single} we show the velocity profiles from our
C$+$D-configuration data (natural weighting) and from single-dish data from the
Effelsberg 100 m radio telescope (Huchtmeier \& Richter 1986). From our
D-configuration data, we find an integrated source flux of 12.9 Jy \kms, which
is similar to three of the four single-dish flux measurements listed in
the Huchtmeier \& Richter catalog (1989). The fourth measurement, the
one from the Effelsberg 100 m radio telescope, is quite a bit higher at
16.9 Jy \kms. The Effelsberg beam at 21 cm is 9\arcmin; a search within
a 9\arcmin\ radius of DDO 43 using NED\footnote{NASA/IPAC Extragalactic
Database; http://nedwww.ipac.caltech.edu/} turned up only a few
continuum sources, so it is unlikely that this discrepancy in the
measured flux is caused by confusion in the Effelsberg beam.

\placefigure{fig:single}

If the Effelsberg flux is correct, we have only recovered 76\% of the
\HI\ flux.  We have checked our flux calibration by comparing the
catalogued flux densities of two (unresolved) continuum sources in the
field with the fluxes measured from our data.  These sources are within
3--4\arcmin\ of DDO 43; they are catalogued in the NRAO VLA Sky Survey
(NVSS\footnote{http://www.cv.nrao.edu/nvss/}; Condon \et\ 1998). In the
C-configuration data, our measured fluxes (44.6 and 94.1 mJy) are an
average of 7\% lower than the NVSS values of 48.9 and 97.9 mJy; for the
D-configuration (42.4 and 88.6 mJy) they are 12\% lower. If we assume
this is accurate and correct our integrated flux for DDO 43 by 12\% for
the D-configuration data, we are still only detecting 85\% of the
reported Effelsberg value.

Another possibility is that the VLA interferometer is missing some of
the flux detected by the Effelsberg single-dish instrument.
Interferometers are ``blind'' to structures larger than an angular size
that depends on the projected baseline lengths between the antennas. The
D configuration can detect structures up to 15\arcmin\ across per spectral
channel before they are resolved out by the interferometer.  If the
discrepancy is due to a low-level envelope too large to be detected by
the VLA, it would have to be approximately twice the extent of the \HI\
that we detected (see \S \ref{sec:radio} below). This seems unlikely, as
does missing faint emission with our sensitive observations. As our
estimate agrees with the other three catalogued measurements, it would
seem that the Effelsberg value is suspect.

\section{Results}

\subsection{Optical}

\subsubsection{Morphology} \label{sec:bar}

\begin{figure}
\epsscale{1.0}
\plotone{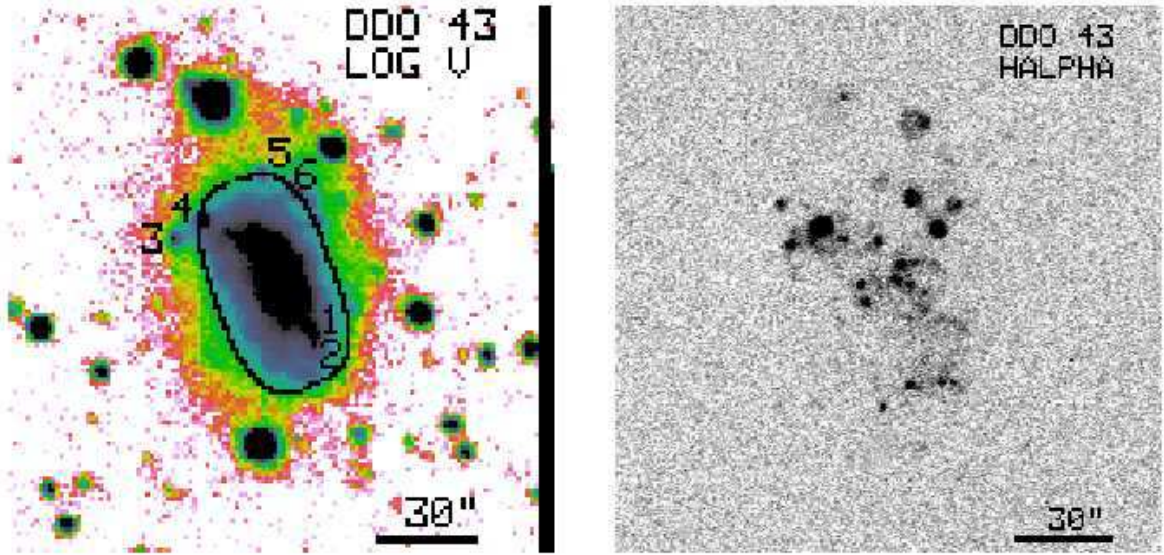}
\caption{Left: False-color display of the
logarithm of our V-band image of DDO 43. We display the image this way
in order to allow immediate comparison of the inner and outer parts of
the galaxy. The smooth elliptical curve superposed in black is the
best-fit bar structure with $c=3.0$ and a semi-major axis of
33\protect\arcsec. Star clusters discussed in the text are numbered. The black
columns along the right edge are from a very bright star to the
southwest in the larger field of view of the CCD that saturated and
bled.  Right: Gray-scale display of our H$\alpha$ image of DDO 43.
Stellar continuum has been removed to leave only nebular emission.  The
field of view and orientation is the same as that of the V-band image.
North is up and East is to the left.
\label{fig:v}}
\end{figure}

The logarithm of our V-band image of DDO 43 is shown in the left panel
of Figure \ref{fig:v}. The most striking feature of the image is the
rather rectangular morphology of the middle isophotes.  This morphology
resembles, for example, that of the Im galaxies DDO 75, NGC 1156, NGC
2366, and NGC 4449. This shape suggests the presence of a stellar bar
potential. The identification of bars in Im galaxies is complicated by
the lack of symmetry provided by spiral arms and is especially difficult
if the bar is comparable in size to the galaxy (Roye \& Hunter 2000).
However, there are some properties that can offer additional evidence
that a bar is present.

First, we fit the rectangular part of the V-band isophotes with a curve
of the form $({{|x|}\over{a}})^c + ({{|y|}\over{b}})^c = 1$, where $a$
and $b$ are the semi-major and minor axes and $c$ is a parameter that
describes the boxiness of the bar structure (Athanassoula \et\
1990). For $c=2$ the structure is an ellipse, for $c<2$ the structure is
``disky,'' and for $c>2$ the structure is ``boxy.'' The best fit for DDO
43 is a curve with $c=3$, a boxy structure centered at 7$^h$ 28$^m$
17.55$^s$, 40\arcdeg\ 46\arcmin\ 09.8\arcsec\ (J2000) with a bar
semi-major axis $a_{bar}$ of 33\arcsec\ at a position angle of
20\arcdeg.  This curve is shown in Figure \ref{fig:v} 
superposed on the logarithm of the V-band image. The Im galaxies NGC
2366 and NGC 4449 
are also best fit with a $c=3$ boxy curve (Hunter, Elmegreen, \& van
Woerden 2001; Hunter, van Woerden, \& Gallagher 1999). Fitting a boxy
elliptical does not require that a bar be present (see discussion of NGC
2366, Hunter \et\ 2001), but it does suggest that one could be present.

\placefigure{fig:v}

A second observational clue to the presence of a bar can be the twisting
of isophotes as one goes from the bar to the disk.  This is clearly seen
in NGC 4449, for example (Hunter \et\ 1999), but is not a necessary
condition. In DDO 43 there is a small shift in the position angle of the
major axis as one goes from the inner isophotes fit with the boxy curve
($\mu_{V_0}=24.4$ mag arcsec$^{-2}$) to outer isophotes.  The inner
rectangle is best fit with a boxy curve with a position angle of
20\arcdeg\ while the outer isophotes have a position angle of
6.5\arcdeg. This is a shift of 13.5\arcdeg\ and it is clearly visible in
Figure \ref{fig:v}.

A third characteristic of a bar is a potential mis-alignment between the
kinematic line of nodes and the major axis of the bar. In a barred
galaxy, the isophotes are not circular in the principal plane, and the
stellar orbits may precess, so the line of nodes 
and morphological axis need not be aligned.  This is the case in, for
example, NGC 1156 (Hunter \et\ 2002).  We will discuss the gas
kinematics in DDO 43 below, but it is appropriate to mention here that
the position angle of the line of nodes of the rotating gas system is
294\arcdeg.  Thus, the kinematic line of nodes and the major axis of the
rectangular part of the galaxy are nearly perpendicular, being
mis-aligned by 86\arcdeg. This is strong evidence that the boxy
appearance of the optical galaxy is in fact due to a stellar bar
potential.

However, if DDO 43 is barred, the bar occupies a large fraction of the
optical galaxy. This is also the case in NGC 1156, NGC 2366, and NGC
4449. In DDO 43 $a_{bar}$/R$_{25}$ is 0.8; where R$_{25}$ is the radius at a
$\mu_{B_0}$ of 25 magnitudes per 
arcsec$^{-2}$. In NGC 2366 this ratio is 1.6
and in NGC 4449 it is 0.6. By contrast, in Sd--Sm spiral galaxies
$a_{bar}$/R$_{25}$ is 0.05--0.3 (Elmegreen \& Elmegreen 1985, Martin
1995, Elmegreen \et\ 1996). Thus, DDO 43 and other barred Im galaxies
are unusual in being dominated optically by the bar if it is present.

\begin{figure}
\plotone{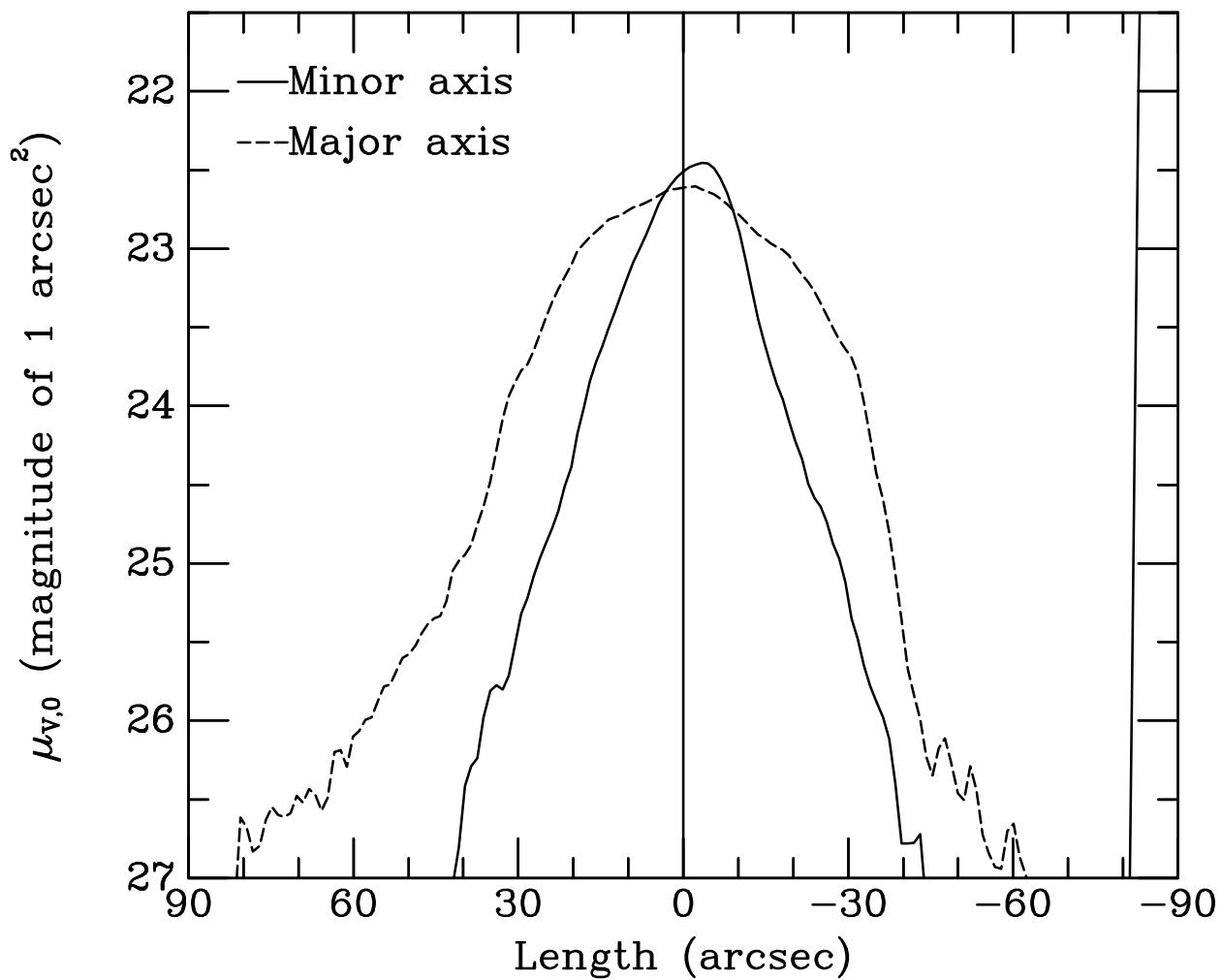}
\caption{Surface brightness cuts along the bar in the
V-band image of DDO 43 (Figure \protect\ref{fig:v}).  Cuts along the
major axis (position angle 20\protect\arcdeg) and along the minor axis are
shown.  Both were constructed by averaging over 20\protect\arcsec.  The
surface brightness is corrected for reddening.  The left side of the
graph refers to the northeast side of the major axis and the southeast
side of the minor axis.  The spike to the right is due to bleeding
columns of a saturated star.
\label{fig:barcuts}}
\end{figure}

In Figure \ref{fig:barcuts} we show surface brightness cuts in V along
the major and minor axis of the bar. The cuts are 20\arcsec\ wide. The
cuts from DDO 43 are more rounded at the top than cuts across the bar of
NGC 4449, but resemble similar cuts made in NGC 2366. The surface
brightness of bars in late-type spirals are more often fit with an
exponential whereas those in early-type spirals are flatter (Combes \&
Elmegreen 1993, Elmegreen \et\ 1996).  The profiles of DDO 43 resemble
those of a few Scd--Sm galaxies with exponential profiles shown by
Elmegreen \et\ (1996).

\placefigure{fig:barcuts}

\subsubsection{Broad-band Surface Photometry}

\begin{figure}
\epsscale{0.75}
\plotone{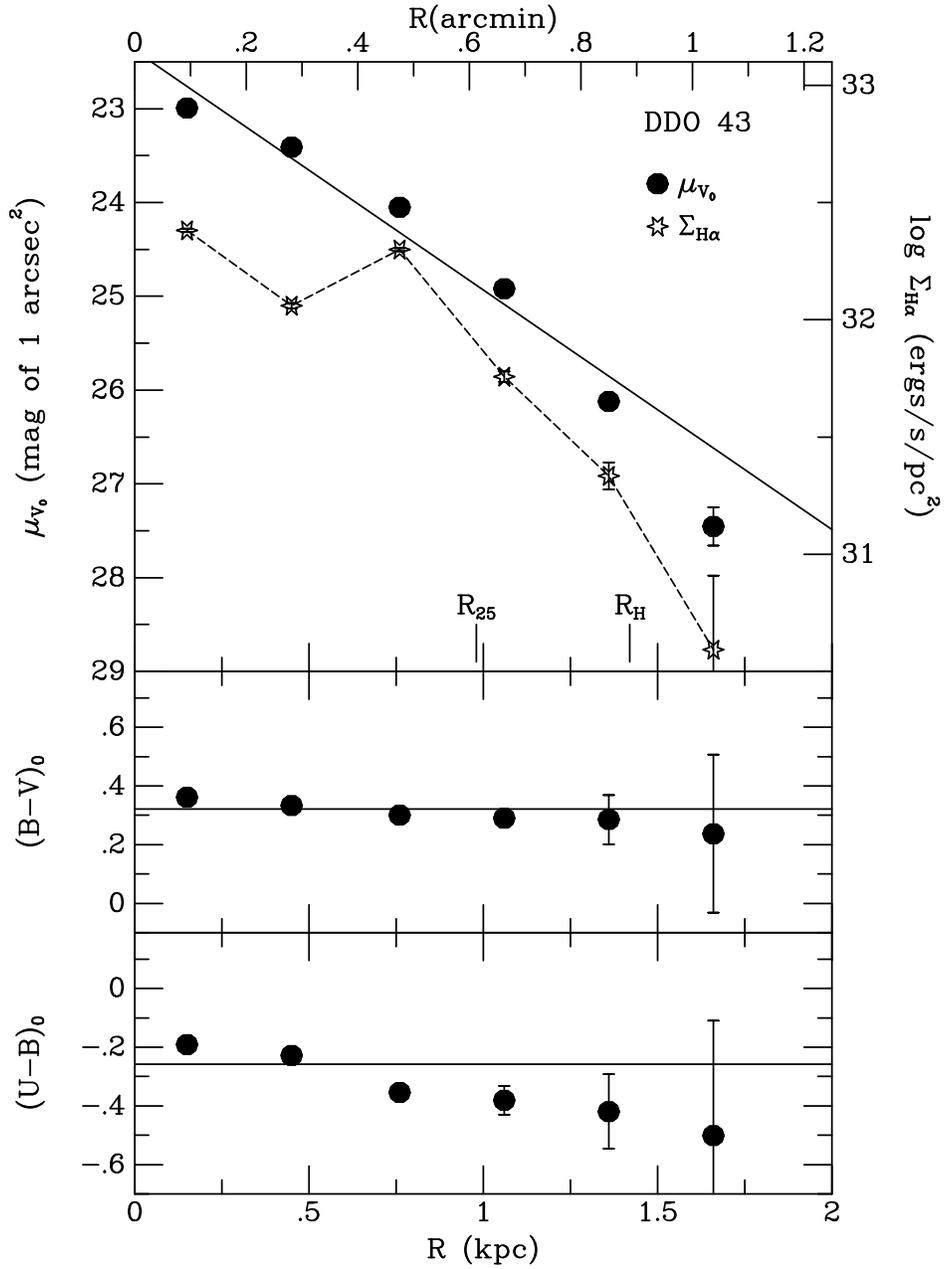}
\caption{UBV and \protect\ha\ azimuthally-averaged
surface photometry of DDO 43 integrated in annuli of
11.3\protect\arcsec\ width.  The UBV photometry is corrected for an
internal reddening of 0.05 mag and a foreground reddening of 0.055
mag. For \protect\ha\ the internal reddening is taken to be 0.1 mag.
The solid line is an exponential fit to the V-band surface brightness
profile.  The scales for $\Sigma_{H\alpha}$, labelled on the right
y-axis, and $\mu_{V_0}$, labelled on the left y-axis, have been set so
that they cover the same logarithmic interval.
\label{fig:ubv}}
\end{figure}

We present UBV surface photometry in Figure \ref{fig:ubv}.  To measure
the surface photometry we used ellipses with a position angle of
6.5\arcdeg, a ratio of minor to major axis ratio $b/a$ of 0.70, and a
semi-major axis step size of 11.3\arcsec.  The $b/a$, position angle,
and center were determined from an outer isophote on a $2\times2$
block-averaged version of the V-band image.  The surface photometry and
colors have been corrected for reddening using a total
E(B$-$V)$_t$=E(B$-$V)$_f$$+$0.05, where the foreground reddening
E(B$-$V)$_f$ is 0.055 (Burstein \& Heiles 1984) and we add 0.05 for
nominal internal reddening. We use the reddening
law of Cardelli, Clayton, \& Mathis (1989) and A$_V$/E(B$-$V)$=$3.1.
Thus, A$_V$ is 0.33 and E(U$-$B) is 0.07.

\placefigure{fig:ubv}

From the reddening-corrected B-band surface photometry $\mu_{B_0}$ we measure
R$_{25}$ to be 37\arcsec, which is 990 pc at the galaxy.  Our
radius is 7\% smaller than that given by RC3.  The Holmberg radius,
R$_H$, originally defined to a photographic surface brightness, is
measured at an equivalent B-band surface brightness $\mu_B = 26.7 -
0.149(B-V)$. For a (B$-$V)$_0$ of 0.32, the Holmberg radius is
determined at a $\mu_{B_0}$ of 26.65 magnitudes of one arcsec$^{-2}$. We
find an R$_H$ of 53\arcsec, which is 1.4 kpc at the galaxy.  Integrated
properties of DDO 43 are listed in Table \ref{tab:int}.

\placetable{tab:int}

We fit an exponential disk to $\mu_{V_0}$, and the fit is shown in
Figure \ref{fig:ubv}.  DDO 43 is fit well with an exponential disk
profile having a central surface brightness $\mu_0$ of 22.4$\pm$0.2
magnitudes of one arcsec$^{-2}$ and a disk scale length R$_D$ of
430$\pm$50 pc. Both these values are normal for Im galaxies (Hunter \&
Elmegreen 2004, de Jong 1996).

DDO 43's integrated colors are typical of Im galaxies (see Figure 2 of
Hunter 1997).  Integrated values are given in Table \ref{tab:int}, and
annular colors are shown in Figure \ref{fig:ubv}.  As in most
irregulars, the colors are, within the uncertainties, constant with
radius.  Two-dimensional color ratio images further substantiate that
there is no large-scale pattern to the colors.

\subsubsection{Star-formation Activity}

We show our \ha\ image of DDO 43 in the right panel of Figure \ref{fig:v}.
There one can see that DDO 43 contains numerous \HII\ regions scattered
across the disk of the galaxy.  The most distant \HII\ region from the
center of the galaxy is found in the northern part of the galaxy just
beyond the Holmberg radius at 1.5 kpc, which is not unusual (see Figure
15 of Hunter \& Elmegreen 2004). The largest \HII\ region, in the
eastern part of the galaxy, has an \ha\ luminosity of $1.0\times10^{38}$
\ergsec, which is equivalent to 10 Orion nebulae (Kennicutt 1984). Thus,
the \HII\ regions in DDO 43 are quite modest in size.

The azimuthally-averaged \ha\ surface brightness is shown with the UBV
surface photometry in Figure \ref{fig:ubv}.  The V-band surface
photometry and \ha\ surface photometry are displayed so that they cover
the same logarithmic interval and, thus, the slopes of the profiles can
be directly compared.  We see that, although star formation is found in
the center of the galaxy, the surface density is less than would be
expected. As a result, there is a depression in the star formation
activity in the central 0.75 kpc of the galaxy. Beyond there the star
formation activity drops off in step with the drop in stellar surface
density, as is common in Im galaxies (Hunter, Elmegreen, \& Baker 1998,
Hunter \& Elmegreen 2004).

The \ha\ luminosity and derived star formation rate of the galaxy are
given in Table \ref{tab:int}. DDO 43 has a star formation rate per unit
area that is in the middle of the range seen in Im galaxies (see Figure
9 of Hunter \&
Elmegreen 2004).  At its current rate of consumption, the galaxy can
turn gas into stars for another 25 Gyr if all of the gas associated with
the galaxy can be used. The timescale to run out of gas is even longer
if recycling of gas from dying stars is efficient.  This timescale to
exhaust the gas is typical of Im galaxies (Hunter \& Elmegreen 2004),
and the galaxy can continue forming stars at its current rate for
several more Hubble times.

\subsubsection{Star Clusters}

\begin{figure}
\epsscale{1.0}
\plotone{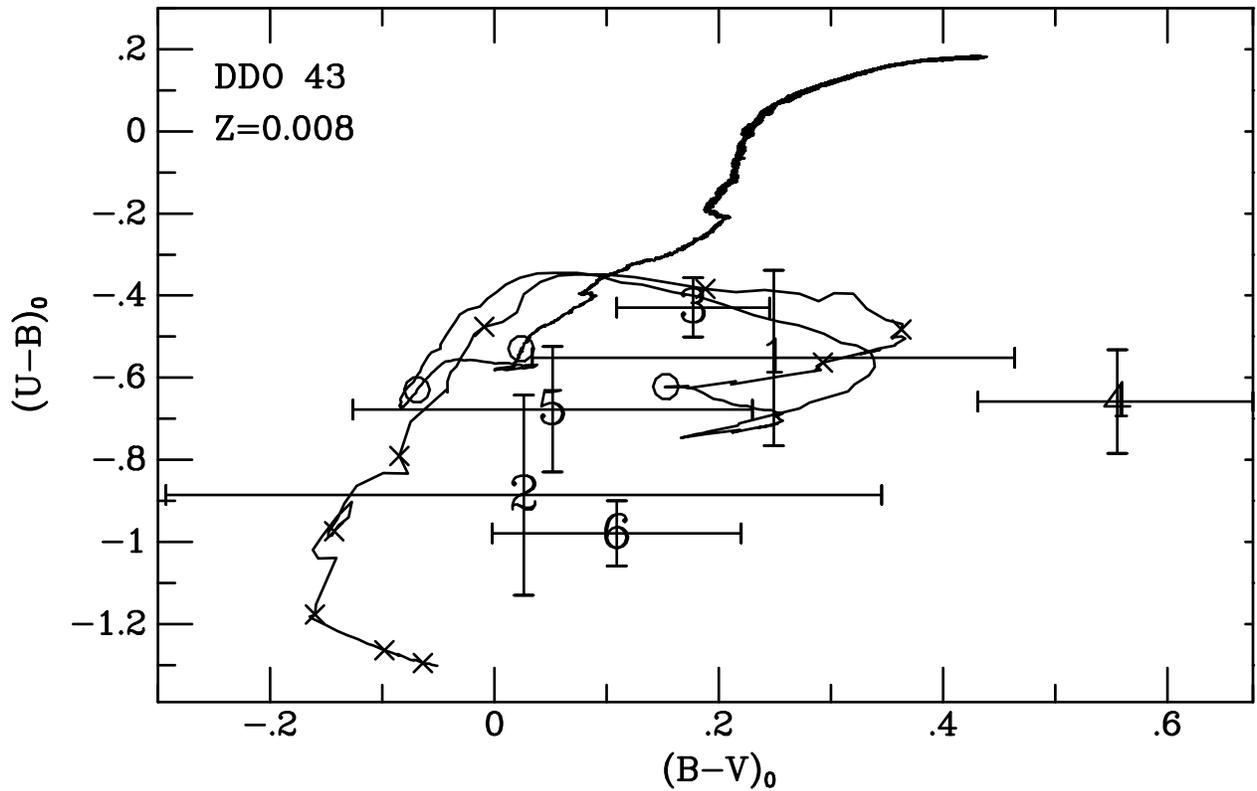}
\caption{UBV color-color diagram with photometry
of the star clusters shown. The numbers of the clusters correspond to
the labels in Figure \protect\ref{fig:v}.  The curved solid line is the
cluster evolutionary track of Leitherer et al.\ (1999) for a metallicity
of Z$=$0.008. The X's on this line mark ages of 1 to 9 Myr in steps of 1
Myr. The open circles on this line mark 10, 20, and 30 Myr time steps.
\label{fig:clus}}
\end{figure}

The V-band image of DDO 43 clearly shows knots that appear to be star
clusters. These are numbered in Figure \ref{fig:v}.  We measured the UBV
photometry of these clusters through apertures that just included the
cluster as seen on the V-band image, radii of 1.7\arcsec\ to
4.0\arcsec. To determine the contribution to the signal from the
underlying galaxy, we used an annulus 4.0--6.2\arcsec. The cluster
photometry is listed in Table \ref{tab:clus} and shown in Figure
\ref{fig:clus} along with the cluster evolutionary track of Leitherer
\et\ (1999) for a metallicity of $Z=0.008$.  The evolutionary track is
used to determine the allowed ages of the clusters. The uncertainties of
the colors allow a range of ages in most cases. However, clusters 3--6
are associated with \ha\ nebular emission, so the younger end of the
range of allowed ages is more likely for these clusters.

\placefigure{fig:clus}
\placetable{tab:clus}

Because star clusters fade and redden as they evolve, to compare the
photometry of clusters requires that one refer them to a common age. We
use the fiducial age of 10 Myrs adopted by Billett, Hunter, \& Elmegreen
(2002).  The M$_V$ that the DDO 43 clusters would have at an age of 10
Myrs (M$_V$(10 Myr)) are given in Table \ref{tab:clus}.  We can then use
M$_V$(10 Myr) as a way to indirectly compare the masses of the clusters.
We adopt the quantitative definitions of ``populous'' and ``super'' star
clusters initiated by Billett \et: M$_V$(10 Myr)$\leq-10.5$ for a super
star cluster and $-9.5\leq$M$_V$(10 Myr)$<-10.5$ for a populous
cluster. Super star clusters are the most luminous and massive clusters;
populous clusters are less so but still substantial.  Only cluster 1 in
DDO 43 could be as luminous as a populous cluster if it is as old as the
upper end of its range of allowed ages.  Since nebular emission is
associated with cluster 4, the lower age and hence lower M$_V$(10 Myr)
is more likely, and so cluster 4 is not likely a populous cluster.
Clusters 2--6 are relatively small.

\subsection{\HI} \label{sec:radio}

\subsubsection{Morphology}

\begin{figure}
\epsscale{0.7}
\plotone{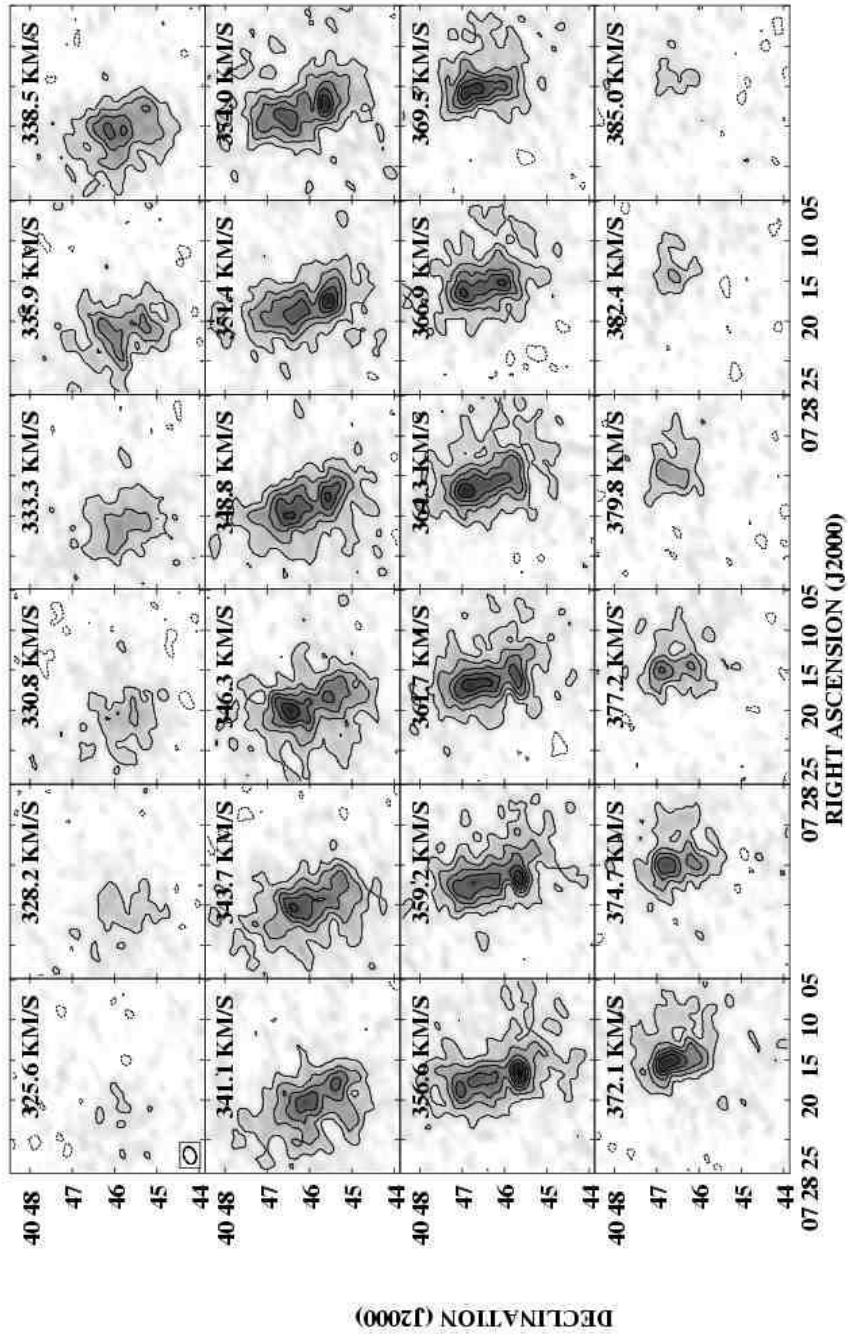}
\caption{Channel maps from the naturally-weighted \protect\cd\
  \protect\HI\ data cube. Each channel is labelled with its heliocentric
  velocity; the beam FWHM (shown in the lower left corner of the first
  panel) is 20.30\protect\arcsec$\times$14.4\protect\arcsec. Contours
  are drawn at -2 (dashed lines), 3, 6, 9, 12, and $15\sigma$\, where
  1$\sigma$ is the measured r.m.s.\ in a signal-free channel and is 1.05
  mJy Beam$^{-1}$ for this data cube.
\label{fig:cdchanmaps}}
\end{figure}

The naturally-weighted \cd\ channel maps are shown in
Figure~\ref{fig:cdchanmaps}. There is emission in the velocity range
from approximately 325 \kms\ to 385 \kms. Rotation is present, but
primarily solid-body, which is not unusual for dwarf galaxies (Swaters
1999). The \HI\ distribution appears clumpy, with high density knots
embedded in the more dense regions of the galaxy.

\placefigure{fig:cdchanmaps}

\begin{figure}
\epsscale{1.0}
\plotone{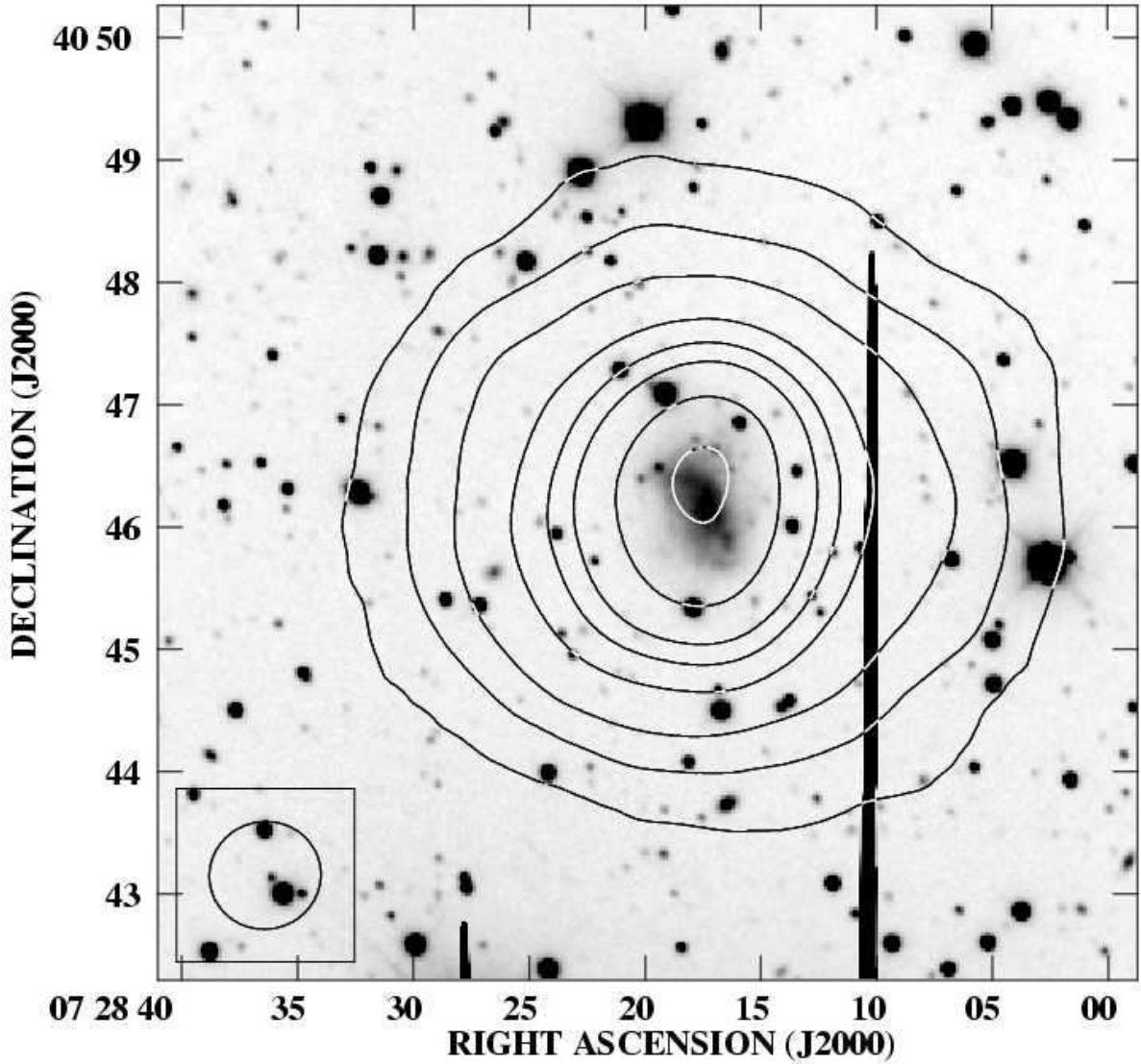}
\caption{Integrated \protect\HI\ flux density contours from the
D-configuration naturally-weighted data cube overlaid on the V
image. The field of view shown is approximately
8\protect\arcmin$\times$8\protect\arcmin. The beam size (FWHM) is shown
in the lower left corner, and measures
54.83\protect\arcsec$\times$52.70\protect\arcsec.  Contours are drawn at
N$_{HI} =$ 1, 5, 10, 20, 40, 60, and 80$\times 10^{19}$
\protect\acm. (1$\times 10^{19}$ \protect\acm\ $=$ 26.17 Jy B$^{-1}$ m
s$^{-1}$.)
\label{fig:donv}}
\end{figure}

The D-configuration alone, with its short baselines and resulting large
beamsize (54.8\arcsec$ \times$ 52.7\arcsec), is unable to resolve any
clear structure in the \HI\ of the galaxy. It does however recover the
most flux and is more sensitive to any large scale structures that may
be present.  The
integrated flux map from the D-configuration data shows only a
relatively circular distribution with the flux density increasing fairly
smoothly towards the center of the galaxy. A total of $9.2\times 10^7$
\msun\ of \HI\ is detected. The \HI\ extends far beyond
the optical body of the galaxy (Figure~\ref{fig:donv}).  The \HI\ radius
at the $5 \times 10^{19}$ \acm\ contour is 146\arcsec\ (3.9 kpc;
corrected for the beam-size), which is 2.8 R$_{H}$.  This is a large
ratio, as indicated by those measured for Sextans A (1.3, Skillman \et\
1988), DDO 50 (1.5, Puche \et\ 1992), and DDO 155 (1.9, Carignan \et\
1990); and by comparison to the sample 
plotted in Figure 13 of Hunter 1997. Although the sample in Hunter 1997
uses the radii at 1$\times 10^{19}$ \acm, DDO 43's ratio using the
smaller 5$\times 10^{19}$ \acm\ contour still falls above the
median. This is also reflected in its high \HI\ mass-to-light ratio 
of 1.64. Although optically DDO 43 is about a magnitude fainter and
about one-half the size of DDO 88, it has a larger \HI\ mass and radius
than DDO 88. \HI\ properties are listed in Table~\ref{tab:int}.

\placefigure{fig:donv}

\begin{figure}
\plotone{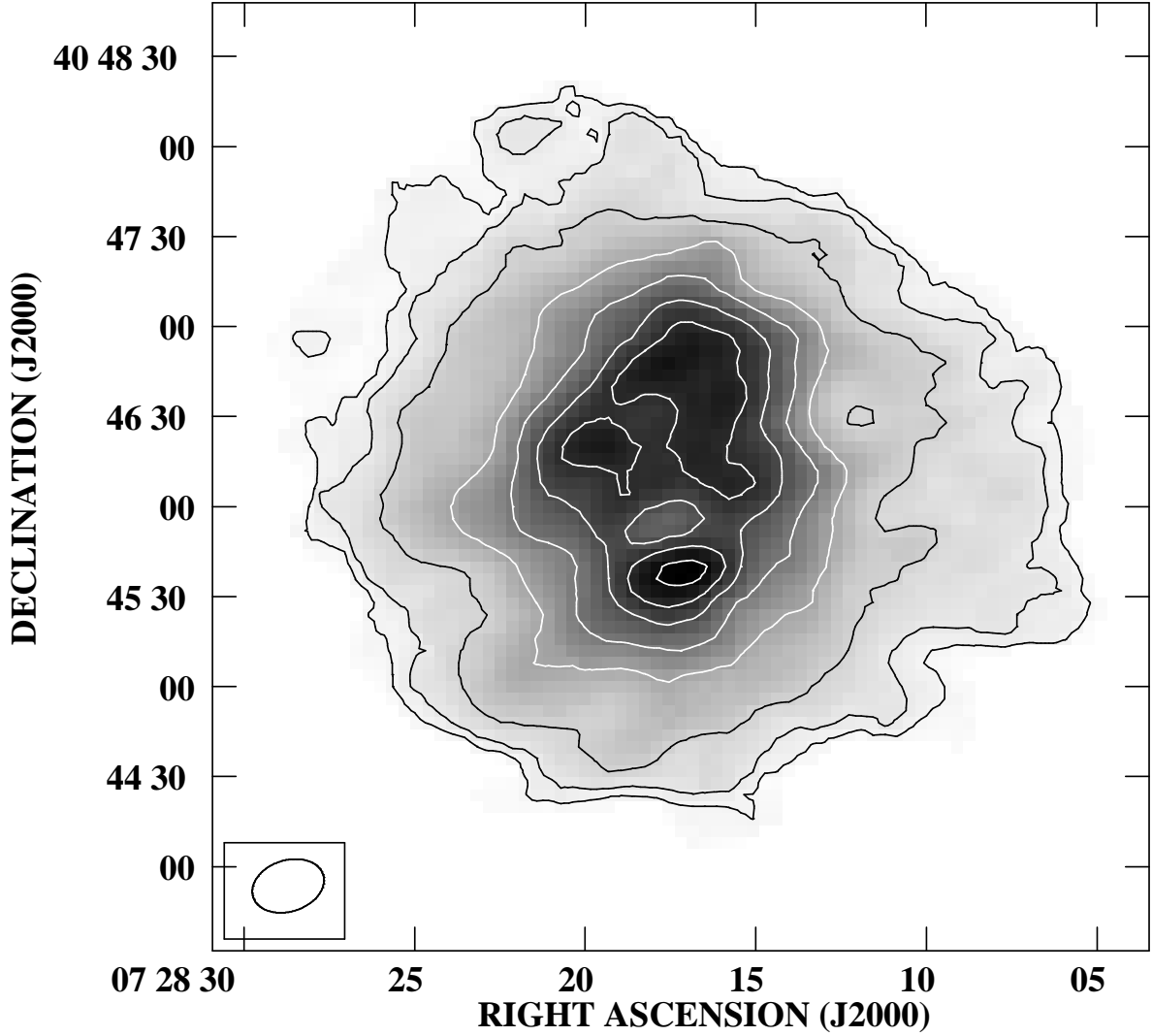}
\caption{Integrated \protect\HI\ flux density from the \protect\cdn\
  cube. The field shown is 5\protect\arcmin$\times$5\protect\arcmin. The
  beam size (FWHM shown in the lower left corner) is
  24.5\protect\arcsec$\times$17.1\protect\arcsec. The contours are at
  N$_{HI} =$0.5, 1, 2, 4, 6, 8, 9.5, and 11 $\times 10^{20}$
  \protect\acm, where 1$\times 10^{19}$ \protect\acm\ $=$ 3.79 mJy
  Beam$^{-1}$ km s$^{-1}$.
\label{fig:cdnm0}}
\end{figure}

\begin{figure}
\plotone{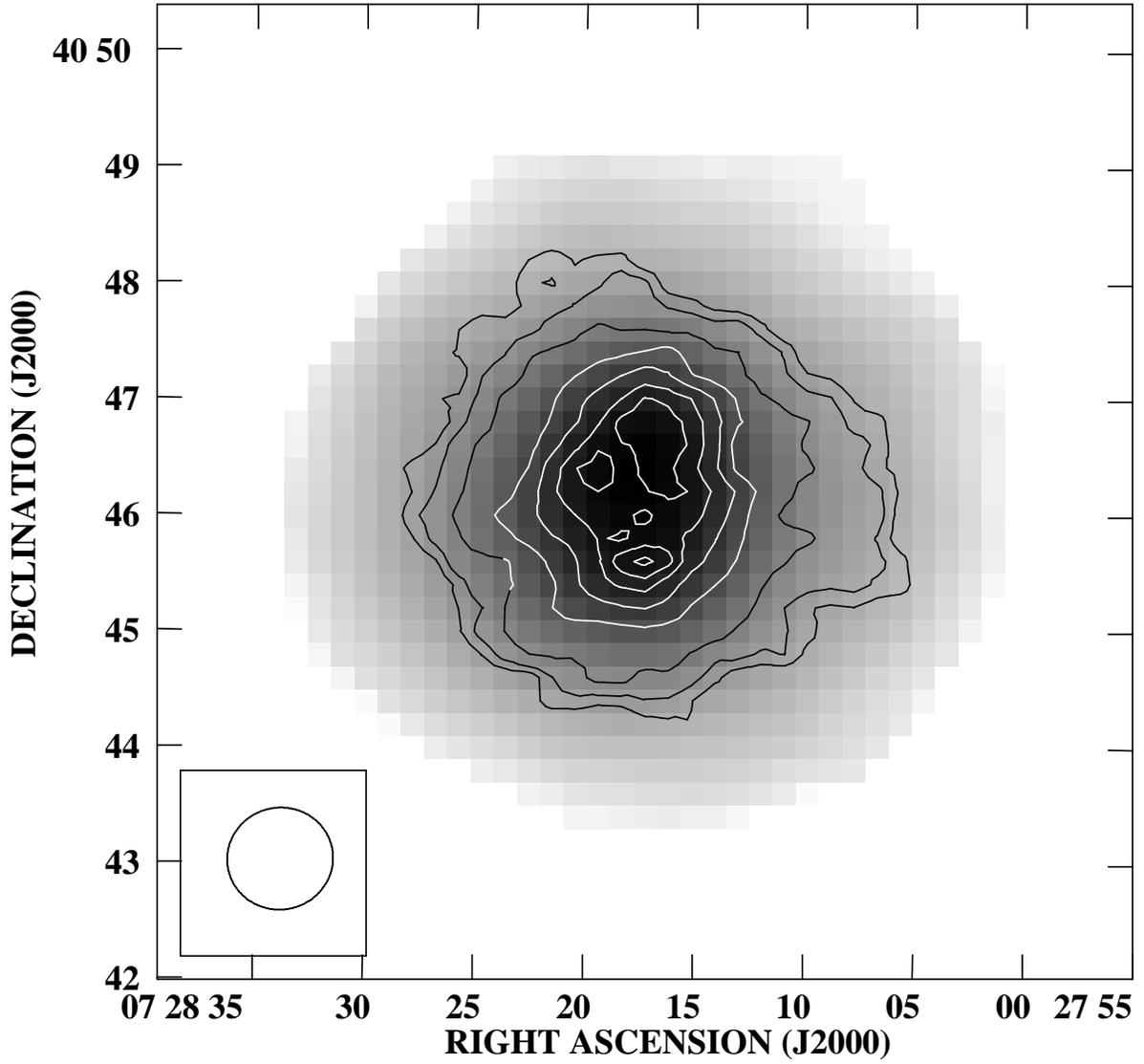}
\caption{D-configuration integrated flux map (greyscale) with contours
  from the \protect\cdn\ data map. The field of view shown is
  approximately 8\protect\arcmin$\times$8\protect\arcmin. Contours are
  drawn at the same levels as in Figure~\protect\ref{fig:cdnm0}.
\label{fig:cdnond}}
\end{figure}

In Figure \ref{fig:cdnm0} we show the integrated \HI\ flux density from
the naturally-weighted \cd\ data. For comparison, the \cd-configuration flux
contours are shown on the D-configuration flux map in
Figure~\ref{fig:cdnond}. Note that a larger region ($8\arcmin \times
8\arcmin$) is plotted for both this figure and Figure~\ref{fig:donv}
than for the others ($5\arcmin 
\times 5\arcmin$). This is necessary to show the full extent of the \HI\
distribution mapped by the D configuration.  From this, we can see
that the extensive low-level \HI\ envelope seen in the D-configuration map is
not detected in the \cdn\ map. The \HI\ in the \cdn\ map can only be
measured out to N$_{HI} = 5\times 10^{19}$ \acm, which is at a radius of
115\arcsec\ (3.1 kpc). Although this extends to only 63\% of the
total D-configuration 
radius, it contains 94\% of the total \HI\ mass measured from the
D-configuration data. Only 6\% is in the outer envelope, so it is indeed
tenuous.

\placefigure{fig:cdnm0}
\placefigure{fig:cdnond}

Although lacking the D configuration's sensitivity, the higher spatial
resolution of the \cdn\ data (FWHM $= 24.5\arcsec\times 17.1\arcsec$),
begins to show some structure in the \HI. There is a small depression
or hole in the gas just south of the galaxy center, with a very high
density knot of \HI\ just south of the depression. There are three more
dense regions just north of the center. The outer contour is not
symmetrical; there is a faint, slightly extended area of emission on the
west side of the \HI\ distribution. The average column density in this
extended region is $\sim 1.1 \times 10^{20}$ \acm\ and the peak column
density is $2.8\times 10^{20}$ \acm.

To examine the small-scale structure features in the \HI\
distribution more clearly and at the highest resolution, we produced an
integrated flux map based only on the 
C-configuration data cube using uniform weighting (beam size
$14.0\arcsec\times 11.0\arcsec = 370\times 290$ pc). Without the larger
number of short baselines provided by the D-configuration, the fainter,
larger-scale structures are not mapped, making it easier to identify the
(brighter) small-scale structures. Not all the information provided by
short baselines is lost however, as the version of the C-configuration
that was used for the observations has the same minimum spacing as the
D-configuration. (This C-configuration, now standard, was originally
known as CS, and has an antenna from the north arm moved to the central
antenna pad).

\begin{figure}
\plotone{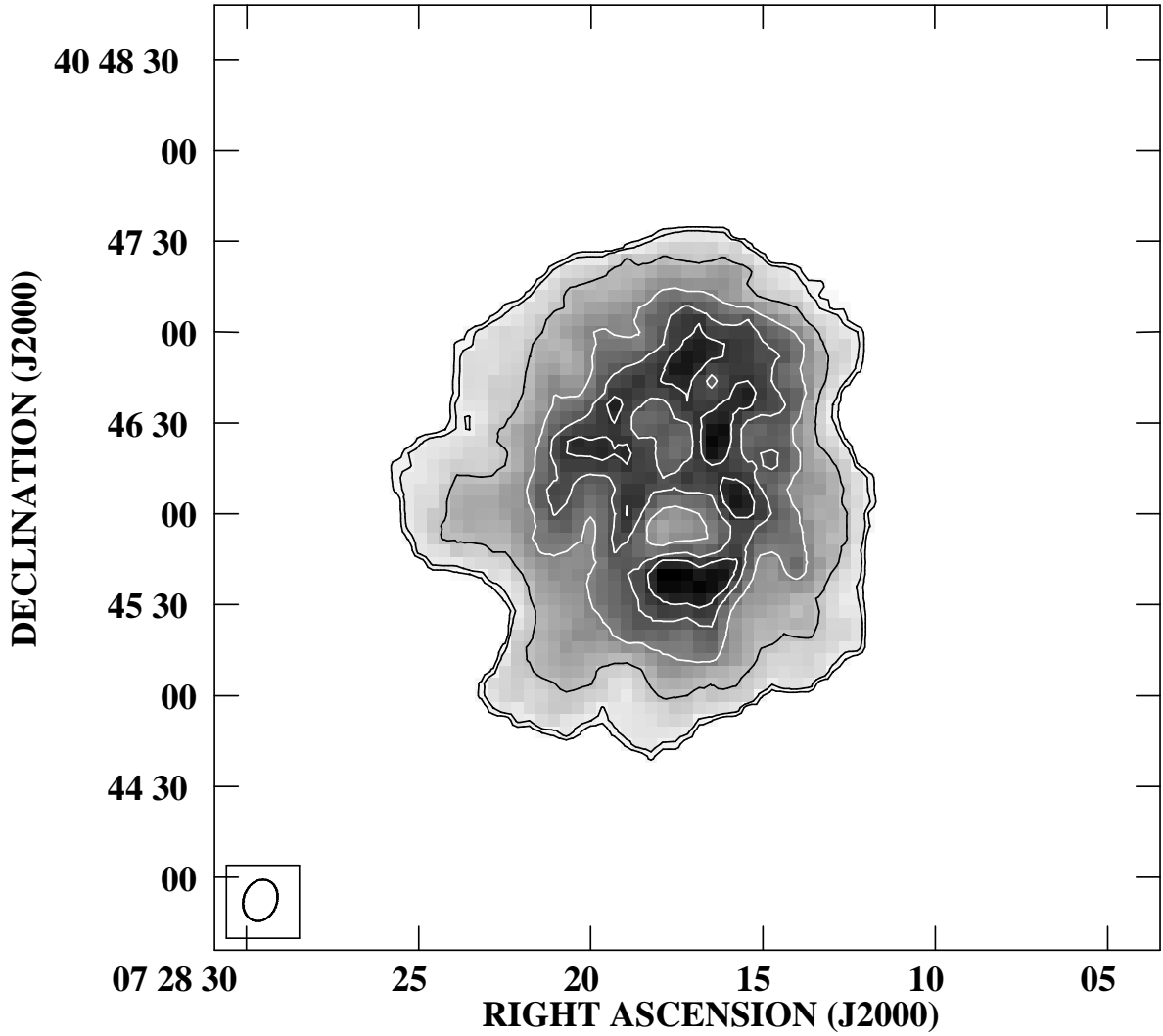}
\caption{Integrated \protect\HI\ flux density from the C configuration
  data (uniform weighting). The field shown is
  5\protect\arcmin$\times$5\protect\arcmin. The beam, indicated in the
  lower left corner, is
  14.0\protect\arcsec$\times$11.0\protect\arcsec. Contour levels are at
  0.5, 1.0, 3.6, 7.2 , 10, and 12$\times 10^{20}$ \protect\acm; 1$\times
  10^{19}$ \protect\acm $=$ 1.395 mJy Beam$^{-1}$ km s$^{-1}$.
\label{fig:cm0}}
\end{figure}

The uniformly-weighted flux map from the C-configuration data more
clearly reveals the high density knots and low density holes in the \HI\
in the inner region of the galaxy (Figure~\ref{fig:cm0}). The large hole
south of center is well-defined in this image, and a second depression
is visible just northwest of center. All these features are located
within the N$_{HI} \geq 1\times 10^{21}$ \acm\ contour, which extends
76\arcsec\ (2.0 kpc) in RA and 100\arcsec\ (2.7 kpc) in DEC. The mass of
the \HI\ inside this high density contour is $4.0\times 10^7$ \msun,
which is 44\% of the total \HI\ mass of the galaxy.

\placefigure{fig:cm0}

We have defined ``knots'' as regions with column densities higher than
$1.2\times 10^{21}$ \acm\ in the C-configuration map. There are six knots
using this criterion; only the brightest and largest (Knot 1), located
in the south of the galaxy, is somewhat resolved. It is approximately
750 $\times$\ 430 pc in size. The average column density of the knots is
$\sim 1.3 \times 10^{21}$ \acm, with a peak column density of $1.48
\times 10^{21}$ \acm\ located in Knot 1. The combined \HI\ mass of the
knots is $\sim 8\times 10^6$ \msun, which is 21\% of the \HI\ mass
contained within the high density region and 9\% of the total \HI\ mass.

``Holes'' are defined as areas with column densities less than 1$\times
10^{21}$ \acm\ that are located inside the high density region
delineated by the N$_{HI} > 1\times 10^{21}$ \acm\ contour. There are
four of these areas: the large, well-defined hole in the south, two
smaller overlapping holes just northwest of center, and a very small
depression to the northeast of that. Only the large hole is resolved; it
is roughly $850\times 530$ pc across, making it larger than Knot 1
just south of it. The average column density in this hole is 7.8$\times
10^{20}$ \acm.

\begin{figure}
\plotone{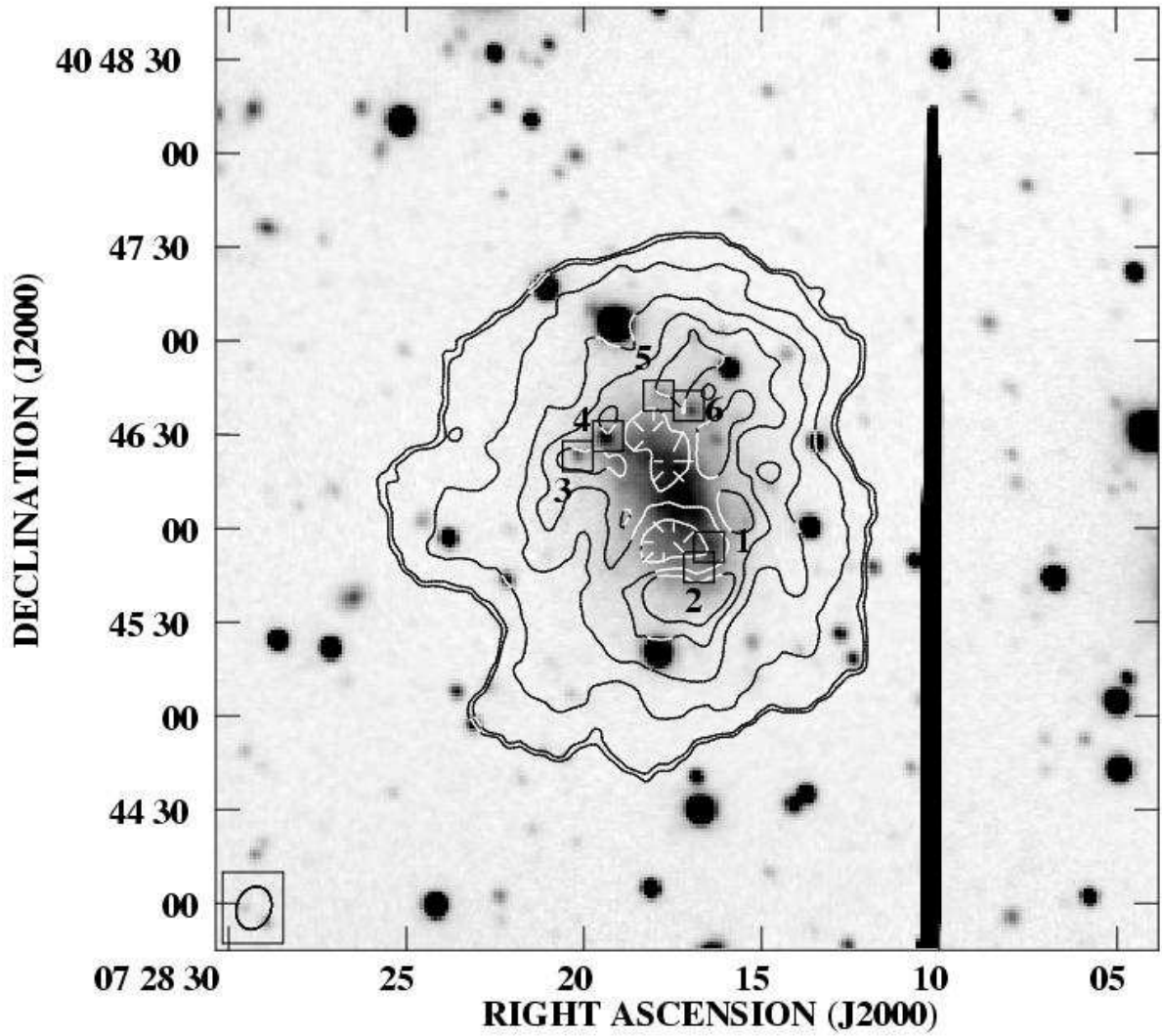}
\caption{C-configuration integrated flux contours on V image. Contours,
  field, and beam-size are the same as in
  Figure~\protect\ref{fig:cm0}. The two most prominent \protect\HI\
  holes are indicated by the contours with the hash marks; the star
  clusters are marked by the boxes and labelled.
\label{fig:cm0onv}}
\end{figure}

The C-configuration flux density contours are shown on the V image in
Figure~\ref{fig:cm0onv}. The higher density \HI\ knots are located in
the outskirts of the optical body of the galaxy, but the high density
region also extends across the center of the bright optical emission in the
galaxy.  The northern edge of the large \HI\ hole just overlaps the
southern part of the bright optical region.

\placefigure{fig:cm0onv}

Of the five star clusters seen on the V image, four are located at the
edges of the high density knots; one (Cluster 1; the largest) is located
just within the western edge of the large hole an another (Cluster 2) is
between the hole and the large \HI\ knot. These clusters are potentially
the two oldest, based on evolutionary models and the lack of any
surrounding \ha\ nebula.

\subsubsection{HI Surface Density}

\begin{figure}
\plotone{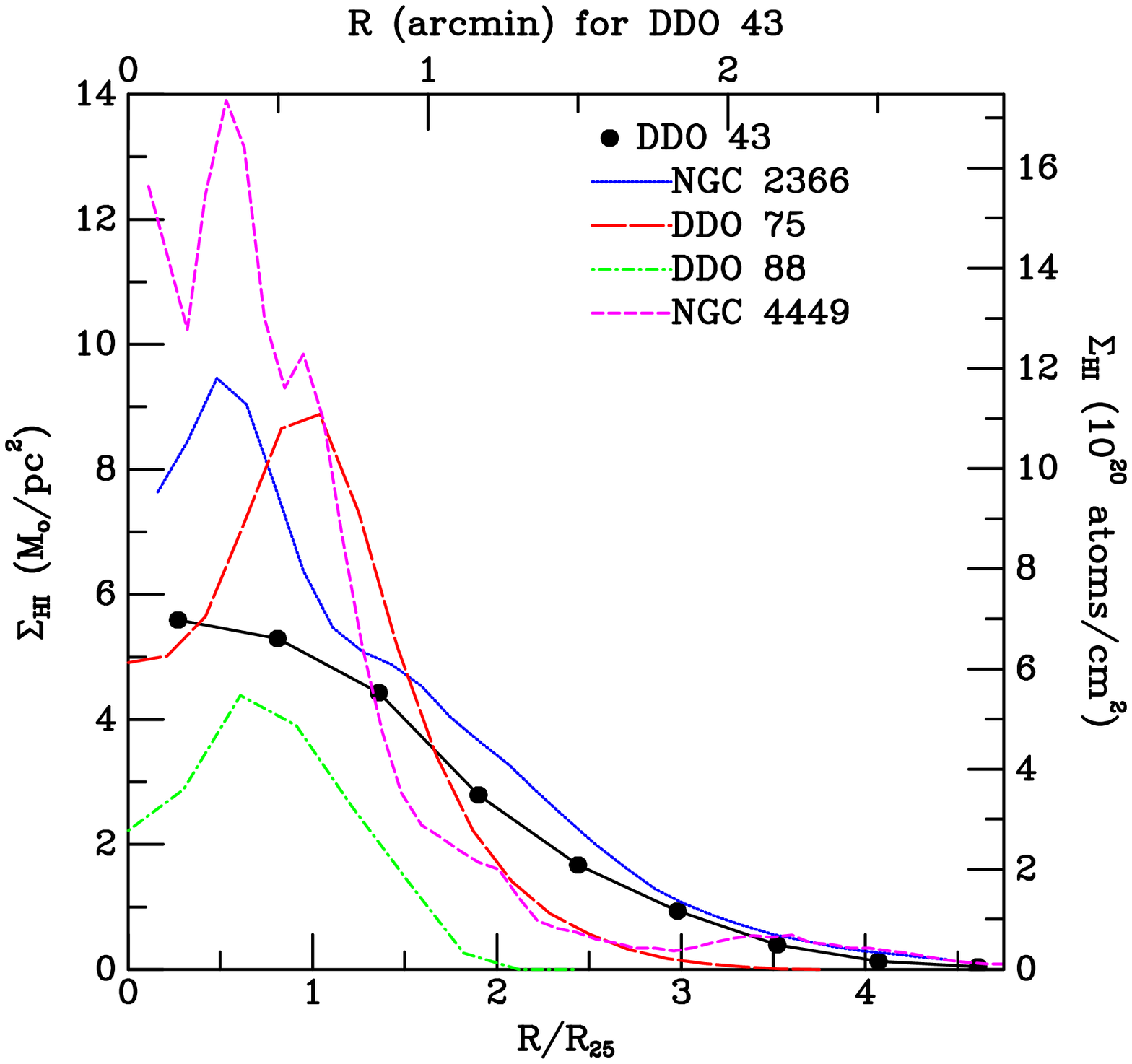}
\caption{Gas surface density in DDO 43 from the \protect\cdn\ data.  The
  \protect\HI\ surface density has been corrected to include He.  For
  comparison we show gas surface density profiles for DDO 75 (Wilcots \&
  Hunter 2002), DDO 88 (Simpson et al.\ 2005), NGC 2366 (Hunter et
  al. 2001), and NGC 4449 (Hunter et al. 1999).
\label{fig:surden}}
\end{figure}

In Figure \ref{fig:surden} we show the azimuthally-averaged surface
density of the \HI\ gas in DDO 43.  We have integrated the \HI\ in
20\arcsec\ radial steps using the naturally-weighted C$+$D-configuration map
(beam size 24.5\arcsec$\times$17.1\arcsec).  We used the center (7$^h$
28$^m$ 17.2$^s$, 40\arcdeg\ 46\arcmin\ 15.2\arcsec; J2000), position
angle (294\arcdeg), and inclination ($i=42$\arcdeg) found from
determining the rotation curve, discussed below. We have multiplied the
\HI\ surface density by 1.34 to include He.

In Figure \ref{fig:surden} we have also plotted several other Im
galaxies, selecting those for which the beam size is comparable to that
of DDO 43. The DDO 43 map had a beam of 650$\times$460 pc; NGC 2366's is
560$\times$480 pc (Hunter \et\ 2001); DDO 75's is 530$\times$440 pc
(Wilcots \& Hunter 2002); and DDO 88's is 600$\times$520 pc (Simpson
\et\ 2005).  We also include NGC 4449, although it was mapped with a
smaller beam (185$\times$150 pc), as an example of a system with large
extended filamentary structure in the outer \HI\ envelope (Hunter \et\
1999).

\placefigure{fig:surden}

We see that averaged in 20\arcsec-wide annuli, the gas surface density
in DDO 43 drops off smoothly with an outer \HI\ envelope that drops off
fairly slowly. Except for the central peak in NGC 2366, DDO 43's \HI\
surface density profile most closely resembles that of NGC 2366.

\subsubsection{Kinematics}

\subsubsubsection{Rotation}

\begin{figure}
\plotone{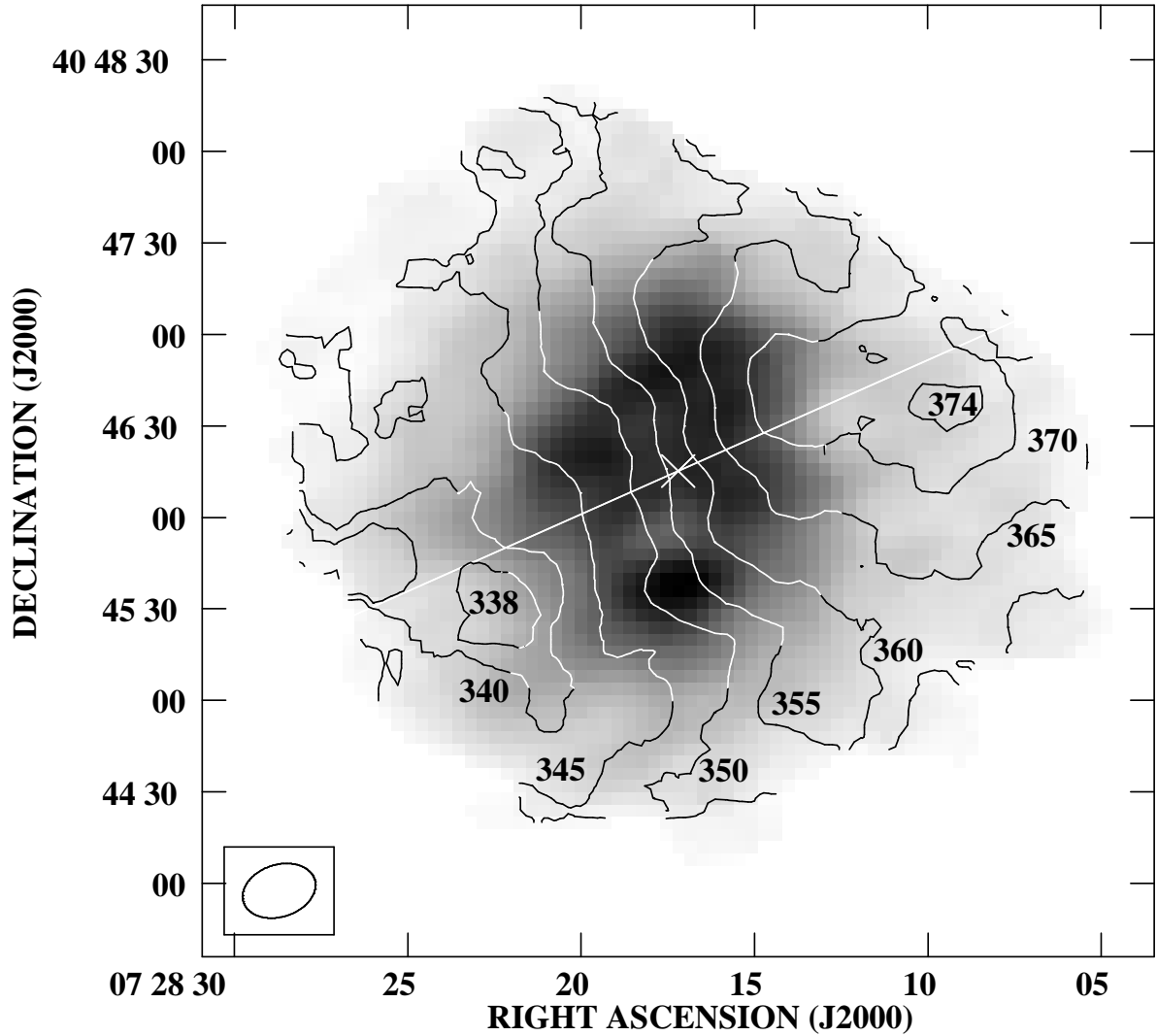}
\caption{Isovelocity contours superposed on an integrated \protect\HI\
  flux map of DDO 43. The moment maps were made from the
  C$+$D-configuration cube made with natural weighting.  Contours are
  marked with velocity values in \protect\kms. The beam size (FWHM shown
  in the lower left corner) is
  24.5\protect\arcsec$\times$17.1\protect\arcsec\ and the velocity
  resolution is 2.6 \protect\kms.  The X marks the kinematic center of
  the galaxy and the straight line the position angle
  (294\protect\arcdeg) that were determined in the fit to the velocity
  field.
\label{fig:vel}}
\end{figure}

The \HI\ velocity field of DDO 43 is shown in Figure \ref{fig:vel} as
contours superposed on the integrated \HI\ map of the galaxy.  One can
see that there is overall rotation with the northwest part of the galaxy
receding and the southeast part approaching.  Superposed on this
rotation, however, is a complex S-distortion that sets in beyond a
radius of about 60\arcsec.  The first kinks in the isovelocity contours
approximately coincide with the \HI\ ridge in the northern part of the
inner galaxy and the large \HI\ complex in the southern part of the
galaxy.  An S-distortion is usually the signature of a warp in the outer
parts of the \HI\ disk. In fact, the velocity field of DDO 43 resembles
that of M83, especially on the approaching side.  M83's velocity field
has been fit with tilted rings that vary in position angle and
inclination (Rogstad, Lockhart, \& Wright 1974).

\placefigure{fig:vel}

\begin{figure}
\plotone{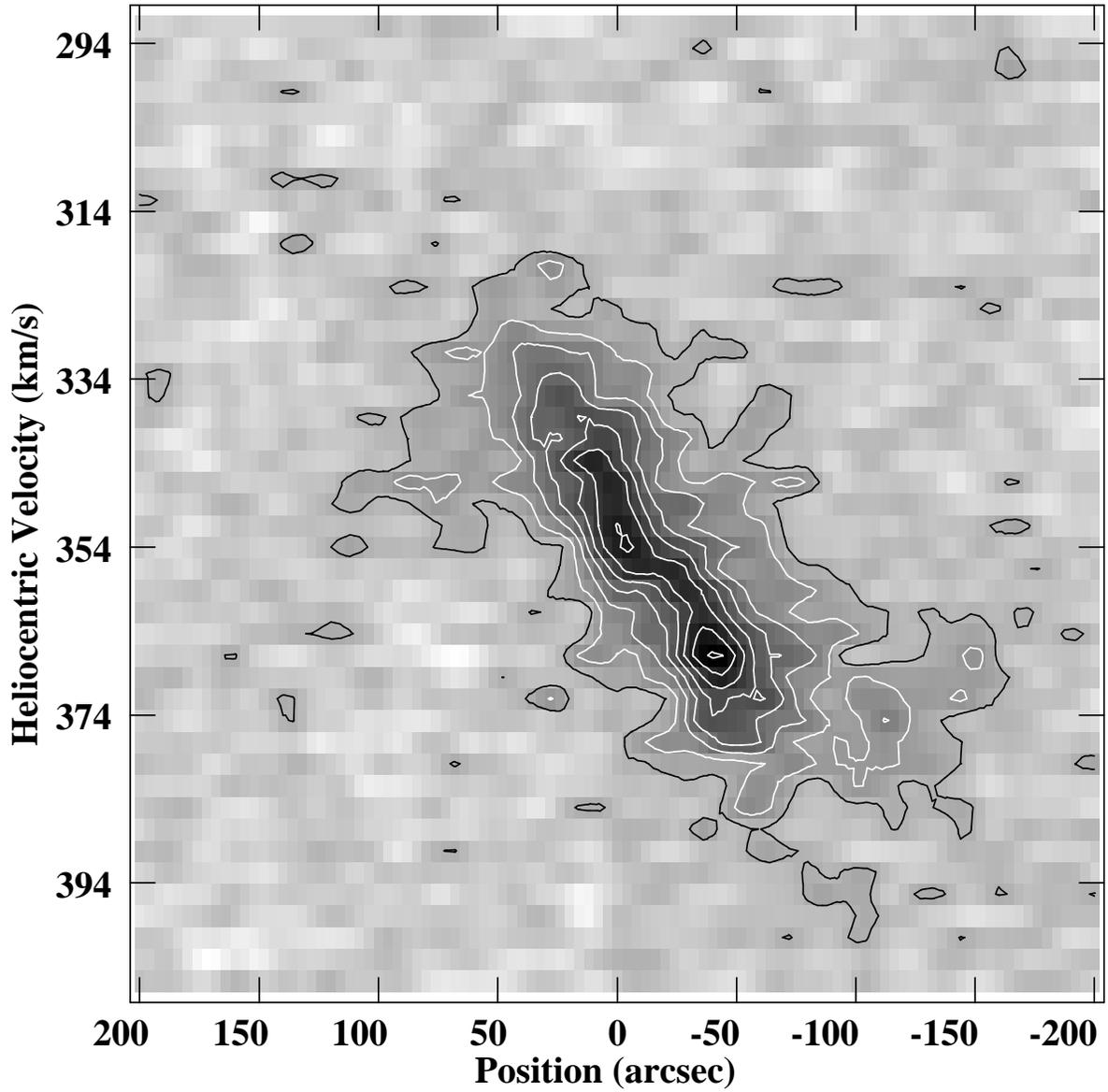}
\caption{Position-velocity cut through the uniformly-weighted
  C$+$D-configuration cube of DDO 43. A cut 20\protect\arcsec\ wide was
  made at a position angle of 294\protect\arcdeg. Contours are at 2, 4,
  6, 8, 10, 12, 14, and 16 $\sigma$; 1$\sigma = 3.7$ mJy/B.
\label{fig:pv}}
\end{figure}

Another view of the velocity field can be obtained from a
position-velocity plot. We have made a 20\arcsec\ cut through the
velocity field of the \cdu\ cube at a position angle of 294\arcdeg, the
kinematic line of nodes that we find below. That position-velocity
diagram is shown in Figure \ref{fig:pv}. One can see there and in Figure
\ref{fig:vel} that the rotation levels off and may turn over in the
outer parts.

\placefigure{fig:pv}

We fit the velocity field determined from the natural-weighted
C$+$D-configuration maps with tilted rings using the task {\sc GAL} in
{\sc AIPS}.  The beam size in this map is
24.5\arcsec$\times$17.1\arcsec, so we used 20\arcsec\ wide rings stepped
every 20\arcsec.  We began by fitting the entire velocity field, the
approaching side only, and the receding side only and allowing all
parameters to vary. Generally speaking the approaching side was harder
to fit than the receding side or both halves together.

\begin{figure}
\epsscale{0.6}
\plotone{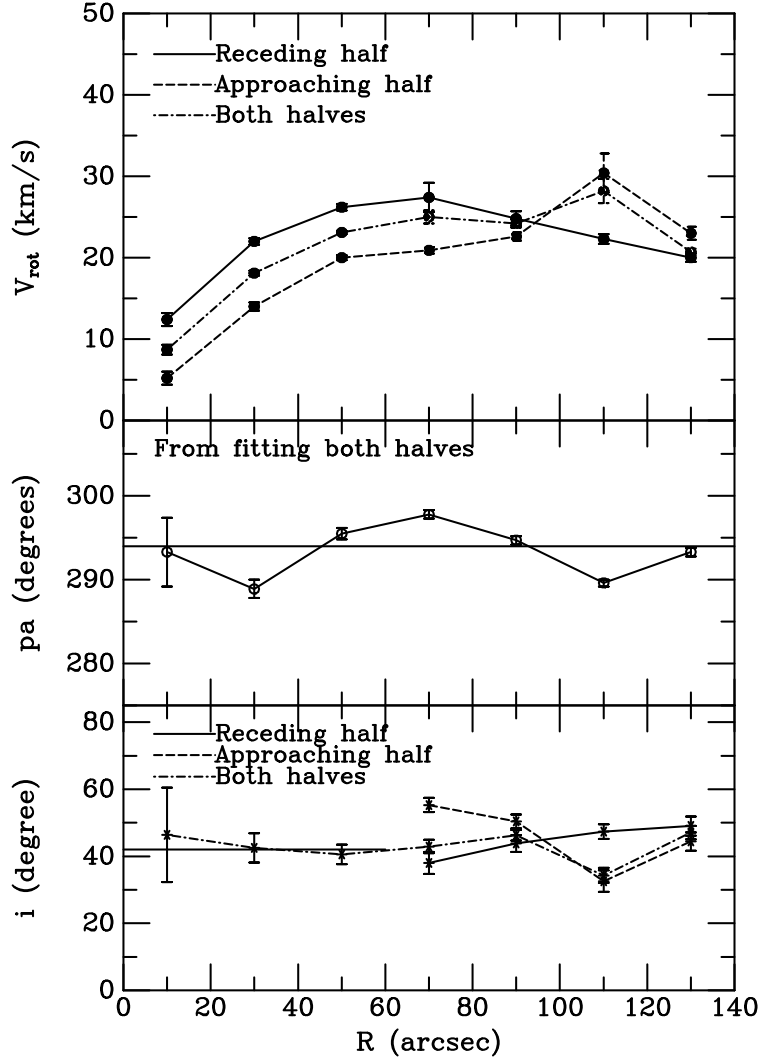}
\caption{Top: Best-fit rotation curve for DDO 43. The rotation curve
  comes from fitting tilted rings of 20\protect\arcsec\ width.  The
  center position, systemic velocity (355 \protect\kms), position angle
  (294\protect\arcdeg), and inclination (42\protect\arcdeg) interior to
  a radius of 60\protect\arcsec\ were fixed in the final fit. The
  inclination beyond 60\protect\arcsec\ was allowed to vary and the
  values of the inclination are shown in the bottom panel.  The
  kinematics of the approaching side of the galaxy are not well fit by
  the model.  Middle: Position angle determined in tilted rings. These
  come from fitting both halves of the galaxy with the center position
  and systemic velocity held fixed.  The solid horizontal line marks the
  average that was then fixed for the rest of the rotation curve
  determination.  Bottom: Variation of inclination in tilted ring
  models. Interior to a radius of 60\protect\arcsec\ the inclination was
  fixed at 42\protect\arcdeg, the average value found from fitting both
  halves of the galaxy.  Beyond 60\protect\arcsec, the inclination was
  allowed to vary in the final determination of the rotation curve.
\label{fig:rot}}
\end{figure}

We found that the center and systemic velocity were quite stable from
ring to ring and among the three solutions, so we determined the average
from the solution to the entire field and fixed those parameters.  The
kinematic center is 7$^h$ 28$^m$ 17.2$^s$$\pm$0.4$^s$, 40\arcdeg\
46\arcmin\ 15.2\arcsec$\pm$4.4\arcsec\ (J2000), and is marked as the
large X in Figure \ref{fig:vel}.  The kinematic center is $-0.35^s$,
$+5.4$\arcsec\ from the center of the optical bar (\S \ref{sec:bar}).
The systemic velocity was found to be 355$\pm$1 \kms. This is the same
as values found by Simpson \& Gottesman (2000; 356$\pm$0.4 \kms) from
VLA observations and by Stil \& Israel (2002; 355$\pm$1 \kms) from WSRT
observations.  We then fixed the center and systemic velocity and fit
the velocity field again. The variation of the position angle with
radius and among the three velocity field fits was small, so we fixed
the position angle at 294$\pm$3\arcdeg. This value for the position
angle is the same as that found by Stil \& Israel (296$\pm$4\arcdeg),
and it is 86\arcdeg\ different from the position angle of the optical
bar.  The average position angle is shown as the solid straight line in
Figure \ref{fig:vel}, and the variation with radius is shown in the
middle panel of Figure \ref{fig:rot}.  We then fixed the position angle
and refit the velocity field.

\placefigure{fig:rot}

The variations in the inclination with the center, systemic velocity,
and position angle fixed are fairly small.  However, if a warp is
present, the inclination need not be the same throughout the
disk. Therefore, we adopted an average inclination of 42$\pm5$\arcdeg\
for the disk interior to a radius of 60\arcsec, where the S-distortion
becomes visible in Figure \ref{fig:vel}.  Beyond 60\arcsec, the
inclination determined for each ring was used in determining the final
rotation speed at that radius.  The variation of inclination with radius
is shown in the bottom panel of Figure \ref{fig:rot}, and the final
rotation curve is shown in the top panel. We show the rotation curve for
fits to the receding and approaching halves separately and for the fit
to the entire velocity field.

The rotation curve in Figure \ref{fig:rot} is relatively normal.  It
rises rapidly in the center and begins to gently turn over around a
radius of 30\arcsec\ (800 pc). By 60\arcsec\ the rotation speed appears
to level off. At a radius of 110\arcsec\ (2.93 kpc) there is an abrupt
rise to a higher rotation speed in the approaching half of the galaxy
clearly caused by a corresponding drop in the inclination angle.
Otherwise, the rotation curve of the receding half drops in the outer
parts.  The maximum rotation speed, leaving out the bump at 110\arcsec,
occurs around a radius of 70\arcsec\ (1.87 kpc) at a speed of 25
\kms. Stil \& Israel (2002) give a rotation speed $V \sin i$ of
17.5$\pm$3.9 \kms\ at a radius of 90\arcsec\ for DDO 43. At that radius
our $V \sin i$ would be the same at 17.6$\pm$0.8 \kms.

\begin{figure}
\epsscale{1.0}
\plotone{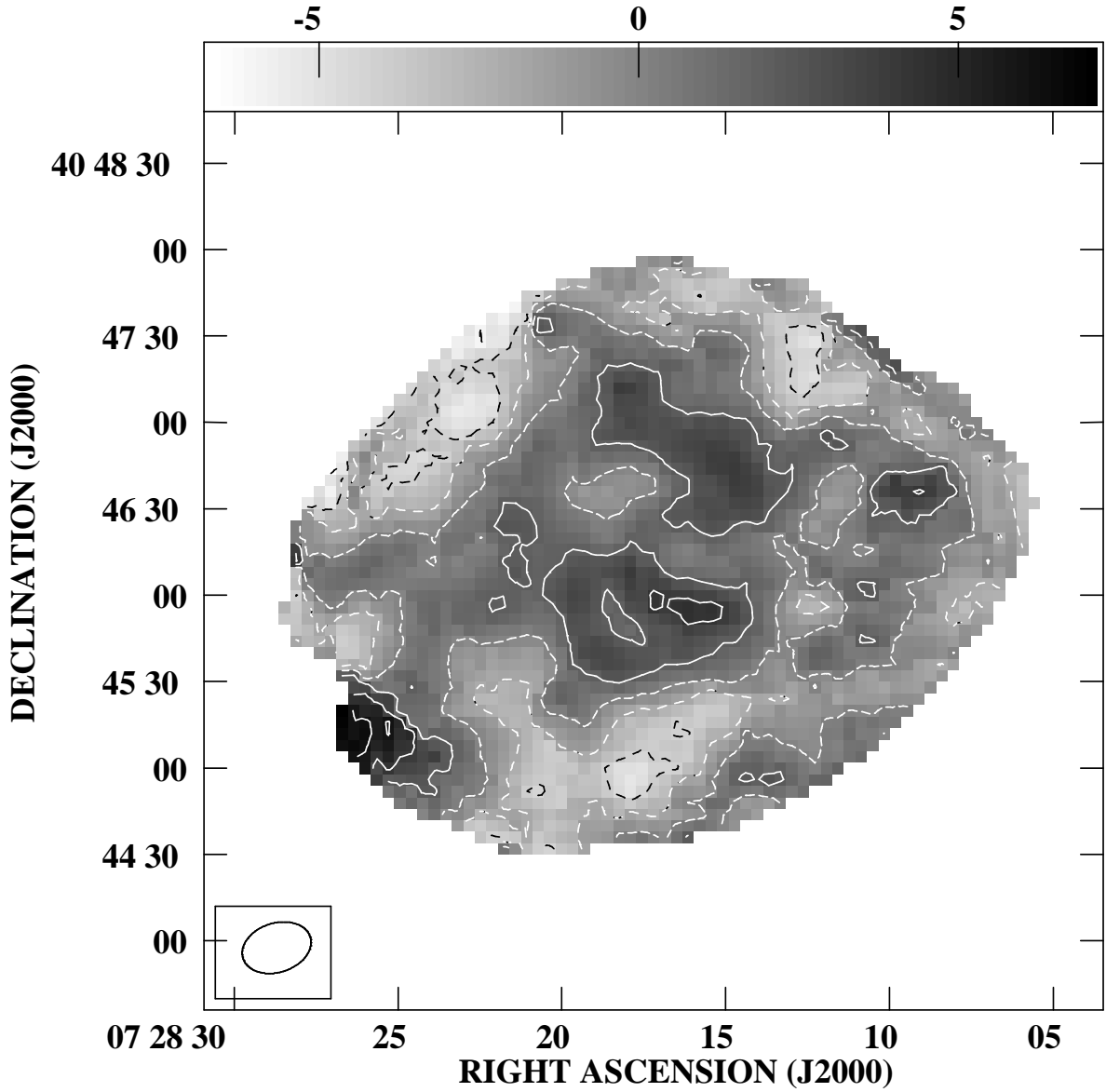}
\caption{Velocity residual map from subtracting the observed velocity
  from the model. Contours are from -6 to +6 \protect\kms\ by 2
  \protect\kms.
\label{fig:resid}}
\end{figure}

\begin{figure}
\plotone{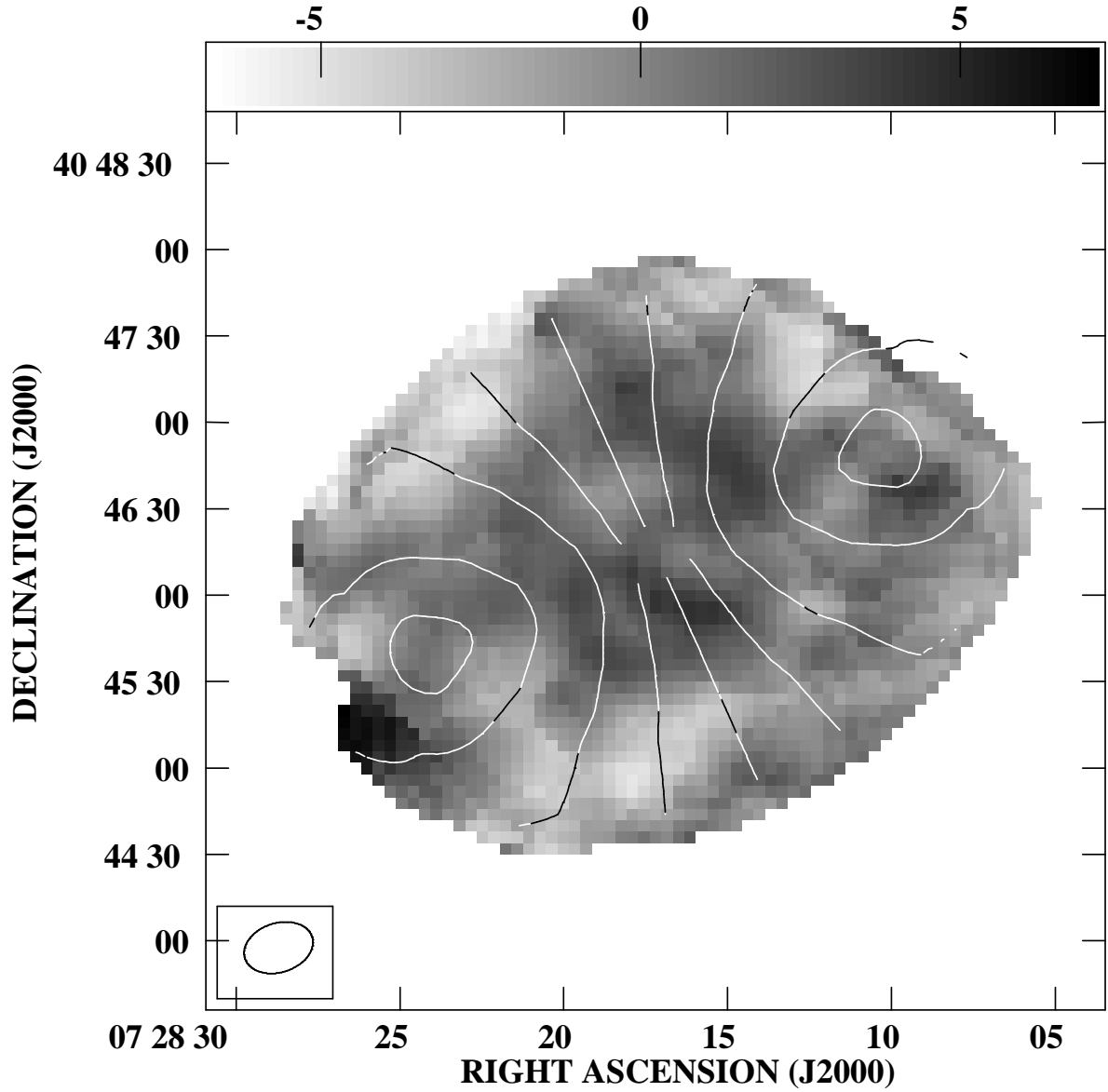}
\caption{Contours from the model velocity
  field plotted on the velocity residual map. Contours are 338, 340,
  345, 350, 355, 260, 365, 370, 372 \protect\kms.
\label{fig:modelonresid}}
\end{figure}

\begin{figure}
\plotone{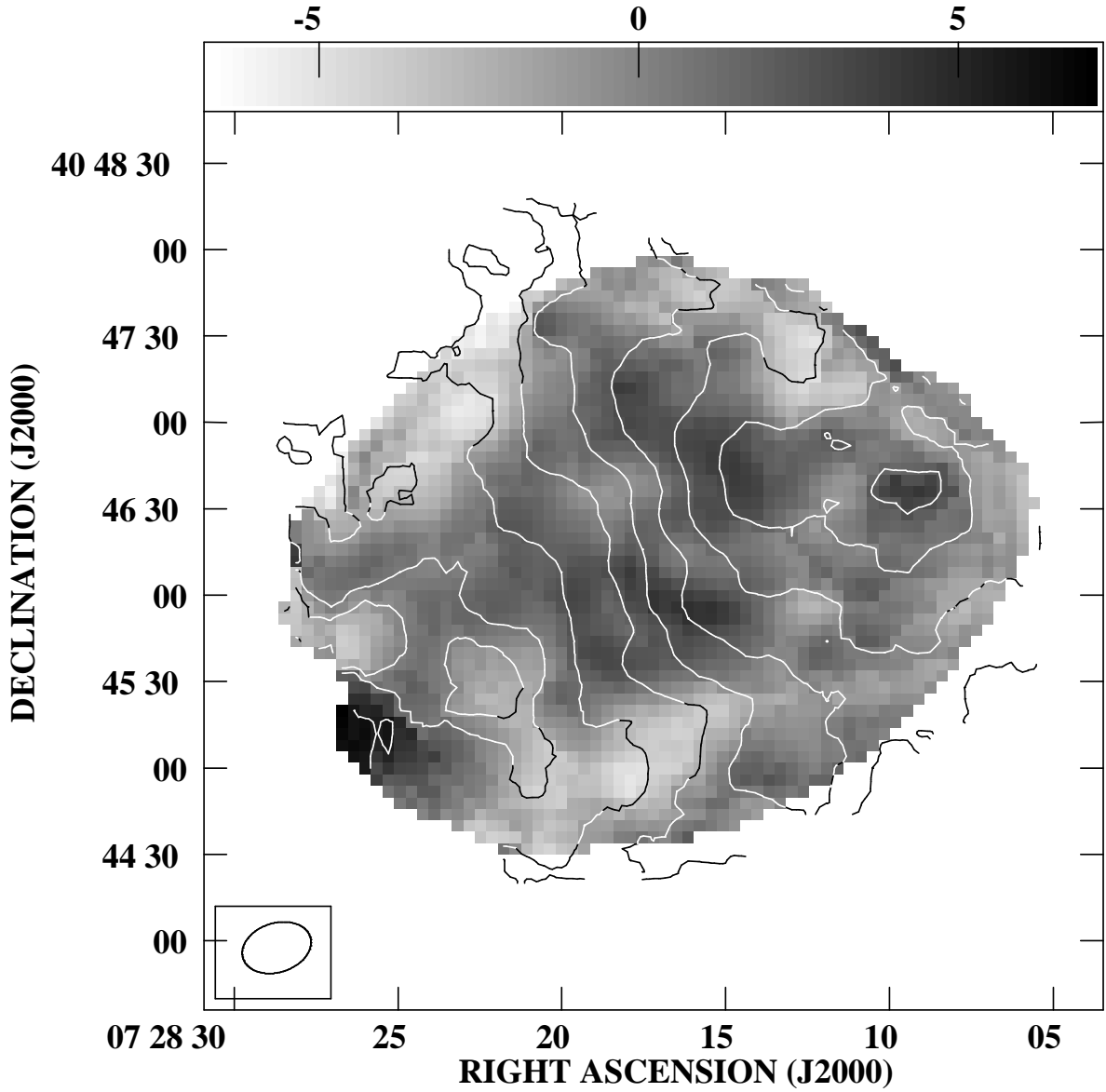}
\caption{The greyscale shows the velocity
  residual map; contours are from the observed velocity field and are at
  338, 340, 345, 350, 355, 360, 365, 370, 374 \protect\kms.
\label{fig:m1onresid}}
\end{figure}

To examine the quality of the fit, we made a model of the velocity
field, then subtracted it from the observed velocity field. The residual
map is shown in Figure \ref{fig:resid}. The values of the residuals
range from -6.6 to 7.1 \kms, so in general, the fit seems quite good. We
have plotted contours from the model velocity field on the residual map
in Figure \ref{fig:modelonresid} and the observed velocity field
contours on the residual map in Figure \ref{fig:m1onresid}. The model
successfully recreates the weak turnover in each side of the velocity
field, but the residual map shows that it had difficulty with the
receding half turnover in the sense that it underestimates it and shifts
it slightly north. There is another region of higher-than-average
residuals just south of the center of the galaxy; this area slightly
overlaps the west sides of the large hole and knot in the \HI\. This is
also near the region of highest velocity dispersion in the galaxy, so
perhaps it is not surprising that the velocity residuals are larger here
as well. From Figure \ref{fig:m1onresid}, it can be seen that most of
the regions of larger residuals (especially negative residuals) coincide
with the kinks in the isovels that may be representative of a warp. The
residuals are still small however, indicating that it's not a
significant warp.

\placefigure{fig:resid}

\placefigure{fig:modelonresid}

\placefigure{fig:m1onresid}

The original listing of DDO 43 as a candidate tidal dwarf by Hunter \et\
(2000) was based on a suggested rotation speed of order 9 \kms. As a
result, on a plot of M$_B$ versus the maximum rotation speed (Figure 1 of
Hunter \et) DDO 43 stood out as unusually luminous for its rotation
speed.  This could imply a deficit of dark matter, a characteristic of
tidal dwarfs (Barnes \& Hernquist 1992).  However, with a maximum
rotation speed of 25 \kms, found here, DDO 43 now lies close to the mean
of the relationship defined by other Im galaxies and spirals in that
plot. Thus, DDO 43 is unlikely to be without dark matter, and unlikely
to be a tidal dwarf.  This is comforting since there was no obvious
nearby post-merger object to have been the parent of DDO 43 if it were a
tidal dwarf.

\subsubsubsection{Velocity Dispersion}

\begin{figure}
\plotone{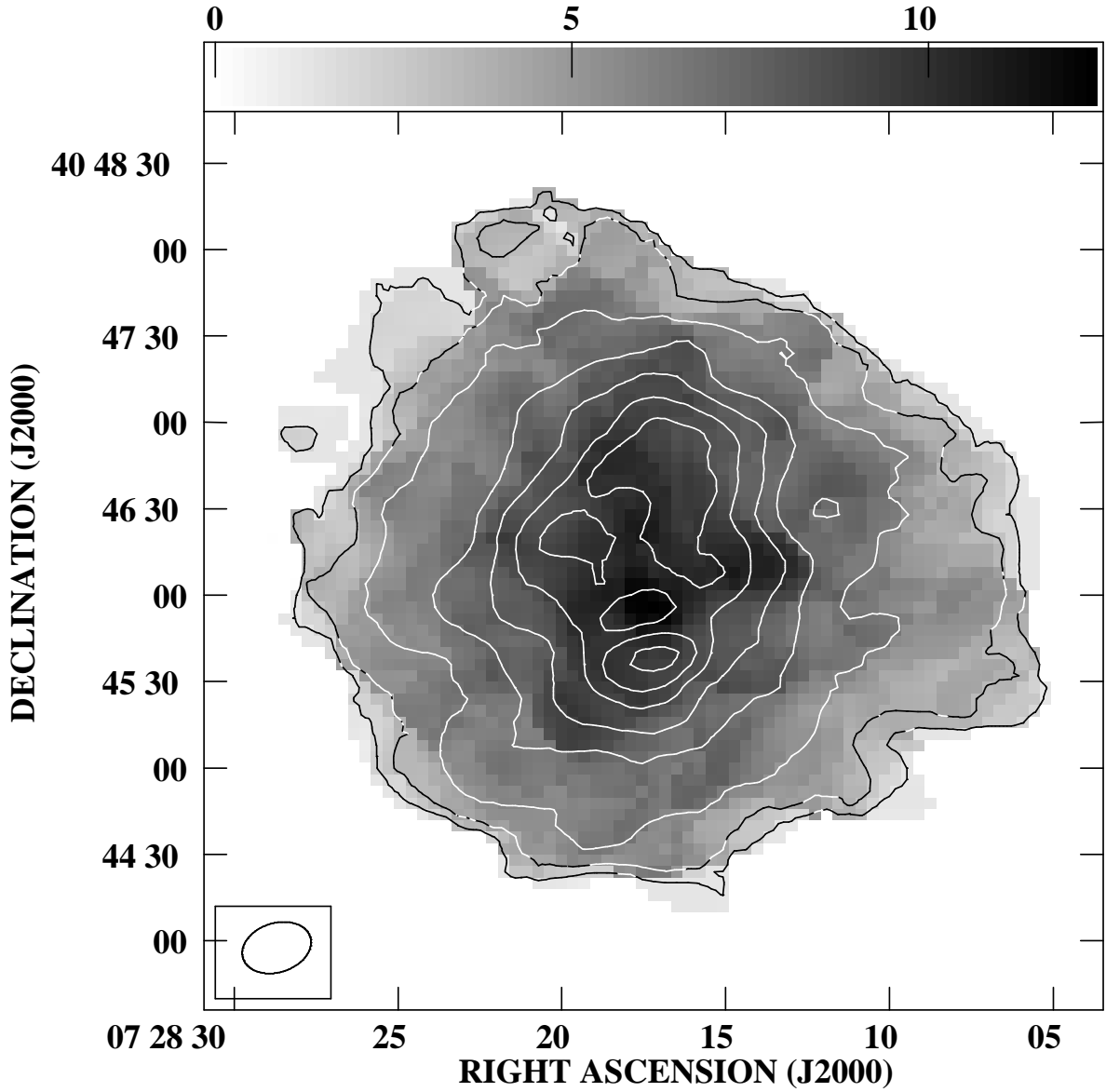}
\caption{The grey-scale shows the velocity dispersion map of DDO 43 made
  from the \protect\cdn\ cube. Contours are of the integrated\protect
  \HI\ map: 0.5, 1, 2, 4, 6, 8, 9.5, and 11$\times 10^{20}$ cm$^{-2}$.
  The contour in the lower left corner is the FWHM of the beam
  (24.5\protect\arcsec$\times$17.1\protect\arcsec).  The bar code across
  the top gives the velocity dispersion in \protect\kms.
\label{fig:veldisp}}
\end{figure}

A grey-scale display of the velocity dispersion map from the \cdn\ data
is shown in Figure~\ref{fig:veldisp} with contours of the integrated
\HI\ superposed.  Within the optical body of the galaxy, the velocity
dispersion is around 10 \kms. This is the value found in most quiescent
gas disks. There are a few spots with higher values up to about 12 \kms,
most notably the large hole. The high column density region is fairly
well demarcated by the 8 \kms\ contour.  Outside of the optical galaxy
the velocity dispersion drops, and values in the extended gas are of
order 5 \kms. Thus, the velocity dispersions in the \HI\ in DDO 43
appear to be quite normal.

\placefigure{fig:veldisp}


However, the second moment map of the galaxy does not tell the whole
story. To better examine the kinematics in the \HI\ knots and holes,
using the \cdu\ cube we plotted the spectra of pixels (averaged across a
beamwidth) for several of the knots and the large hole. The spectra of
the knots are complex. Many exhibit a 
central double peak, often with smaller peaks, sometimes on both sides
of the bright peak, sometimes only visible as broad wings on one or both
sides.

\begin{figure}
\plotone{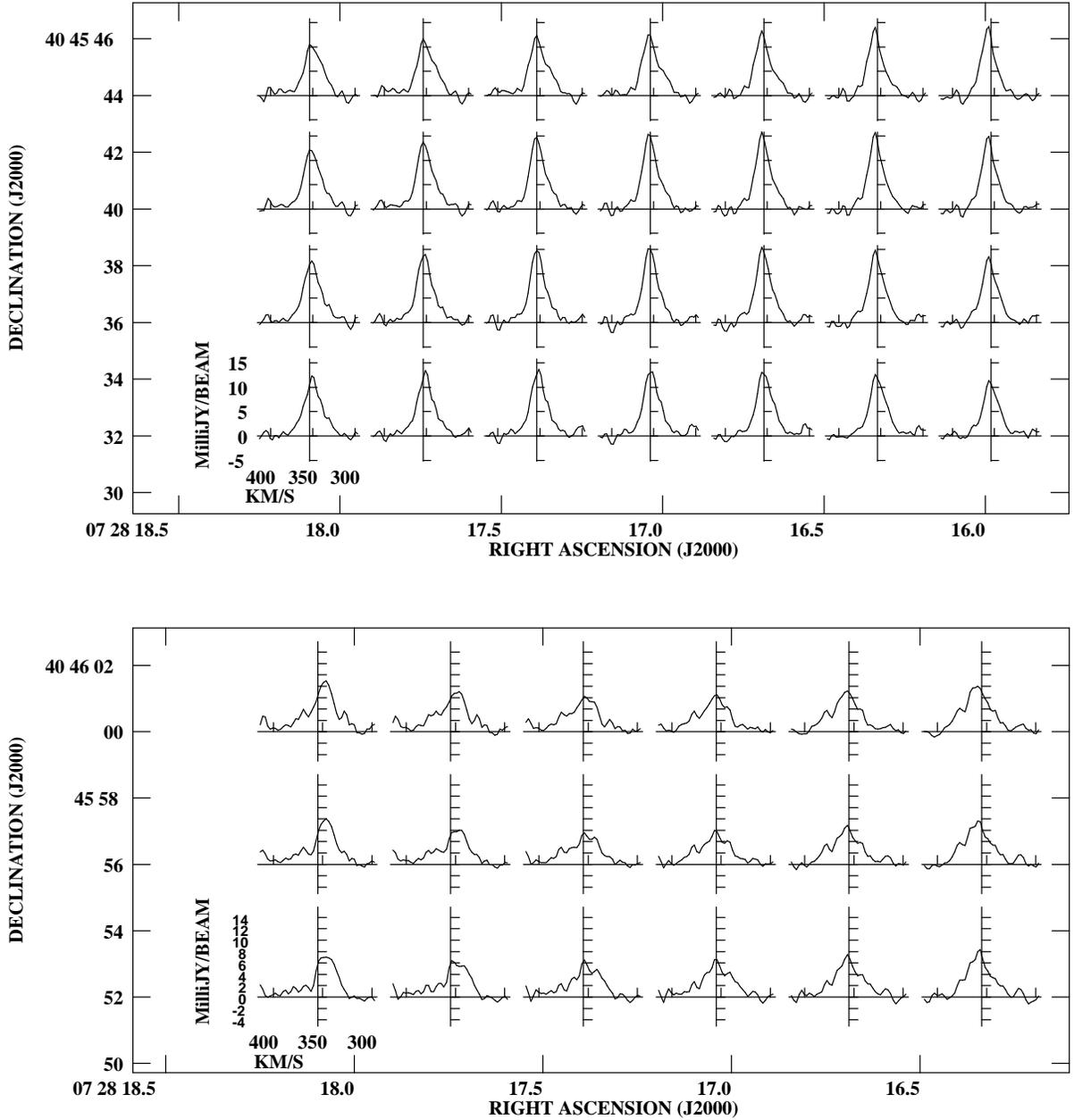}
\caption{Top: Spectra of Knot 1 through the \protect\cdu\
  Hanning-smoothed cube. Each spectrum is from an approximately
  beam-sized area estimated by averaging 4 pixels together. Bottom:
  Spectra of Hole 1 through the \protect\cdu\ Hanning-smoothed
  cube. Each spectrum is from an approximately beam-sized area estimated
  by averaging 4 pixels together.
\label{fig:spectra}}
\end{figure}

We fit gaussians to the Hanning-smoothed beam-averaged spectra of the
brightest knot and the large hole; the spectra are shown in
Figure~\ref{fig:spectra}. For the knot, we were able
to successfully fit three components to 19 of the 28 spectra and two
components to six spectra. Three spectra had large uncertainties to the
fits and so were not included in our analysis. For the 19 spectra with
good fits to three components, the average difference in velocity
between the central and two side components is $\sim \pm 15$\ \kms. The
average amplitudes are 2, 11, and 4 mJy/B going from high to low
velocity. The central and low velocity components are broad, with an
average width of $\sim 17$ \kms. The high velocity component is
narrower, with an average width of roughly half that.

The beam-averaged spectra of the large hole are more complex than those
of the knots, with most being fit by four components. Three of the 18
spectra were fit with three components, and one was fit with
two. Labelling the components 1--4 from high to low central velocities,
components 1 and 3 have average widths of approximately 10 \kms, while
components 2 and 4 have average widths of around 16 \kms. The average
amplitudes are low as expected for a depression in the gas, ranging from
2.2 (component 1) to 5.2 mJy/B (component 2). The components
are separated by 18, 14, and 13 \kms\ on average. The high dispersions
associated with this region are reflected in these spectra; the gas
seems to be somewhat churned up.

\placefigure{fig:spectra}

With these complex spectra, we do not believe that we detect expansion
in the hole. The hole itself is not clearly identifiable in the channel
maps, nor is there any feature in the position-velocity diagrams at the
location and velocity ($\sim$ 350-360 \kms) of the hole that would
indicate expansion or even blowout (e.g.\ Walter \& Brinks 1999). 

\section{Discussion}

\subsection{Current Star Formation and its Relationship to the Gas}

\begin{figure}
\plotone{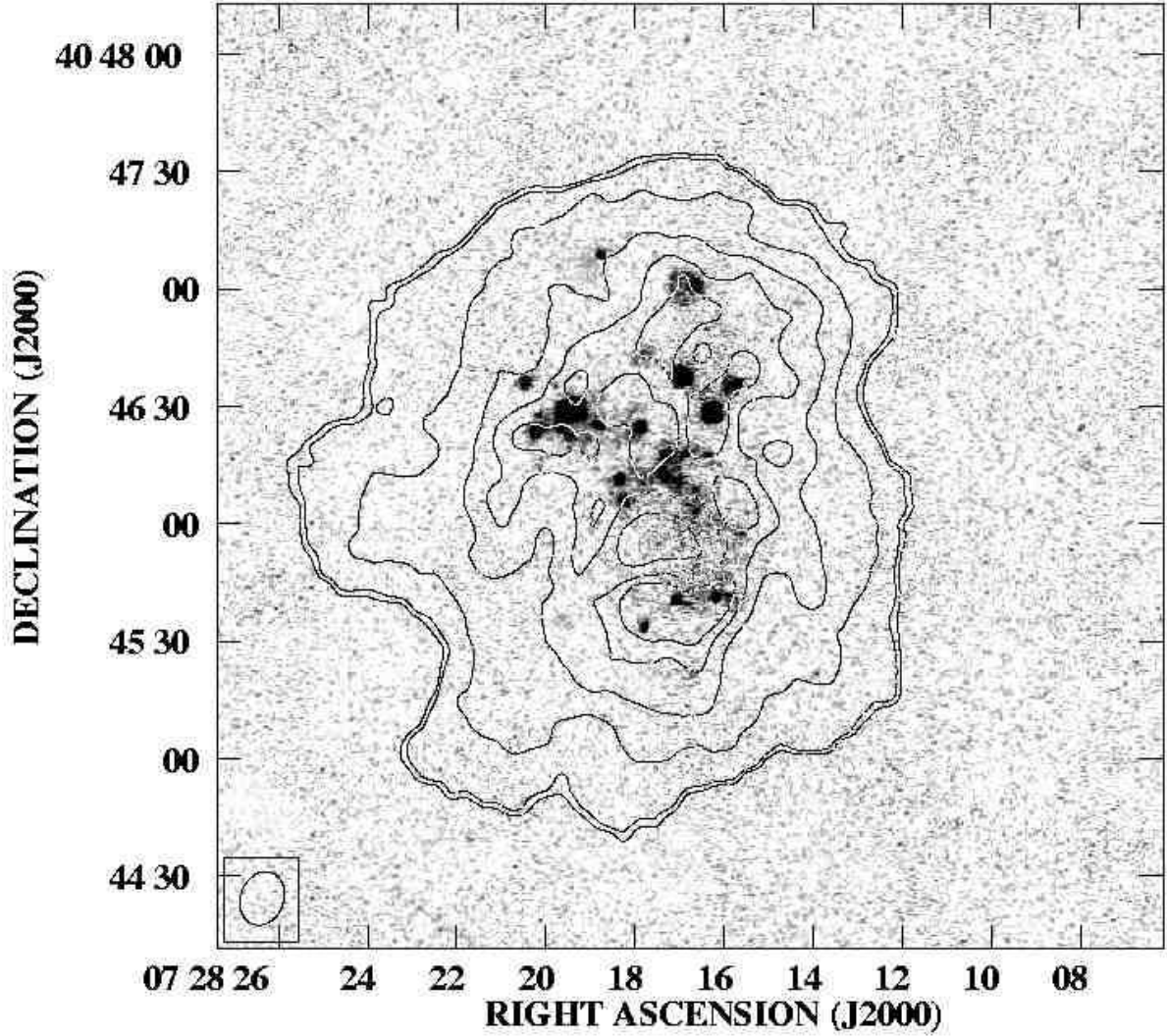}
\caption{Contours of the C-configuration integrated\protect\HI\ map
  shown superposed on our \protect\ha\ image of DDO 43.  The
  \protect\HI\ contours are 0.5, 1.0, 3.6, 7.2, 10, and 12 $\times$
  10$^{20}$ cm$^{-2}$.  The contour in the lower left corner is the FWHM
  of the beam of the C-configuration \protect\HI\ map:
  14\protect\arcsec$\times$11\protect\arcsec.
\label{fig:conha}}
\end{figure}

In Figure \ref{fig:conha} we show contours of \HI\ from the C-configuration map
superposed on our \ha\ image. We see that many of the \HII\ regions,
including the brightest ones, are located on a ridge of \HI\ and in an
\HI\ complex to the south that together appear to form a partial ring
around the center of the galaxy.  These \HII\ regions are found within
an observed \HI\ column density exceeding 1$\times10^{21}$ \coldens.  In
addition, the \HII\ regions are primarily associated with local peaks in
the \HI. This association of star-forming regions with local \HI\ peaks
with observed column densities higher than 10$^{21}$ \coldens\ is often
seen in Im galaxies. Hunter \et\ (2001) argue that the formation of gas
clouds in Im galaxies may be aided by the minimal levels of shear in
these systems.

\placefigure{fig:conha}

\begin{figure}
\includegraphics[scale=0.8,angle=-90]{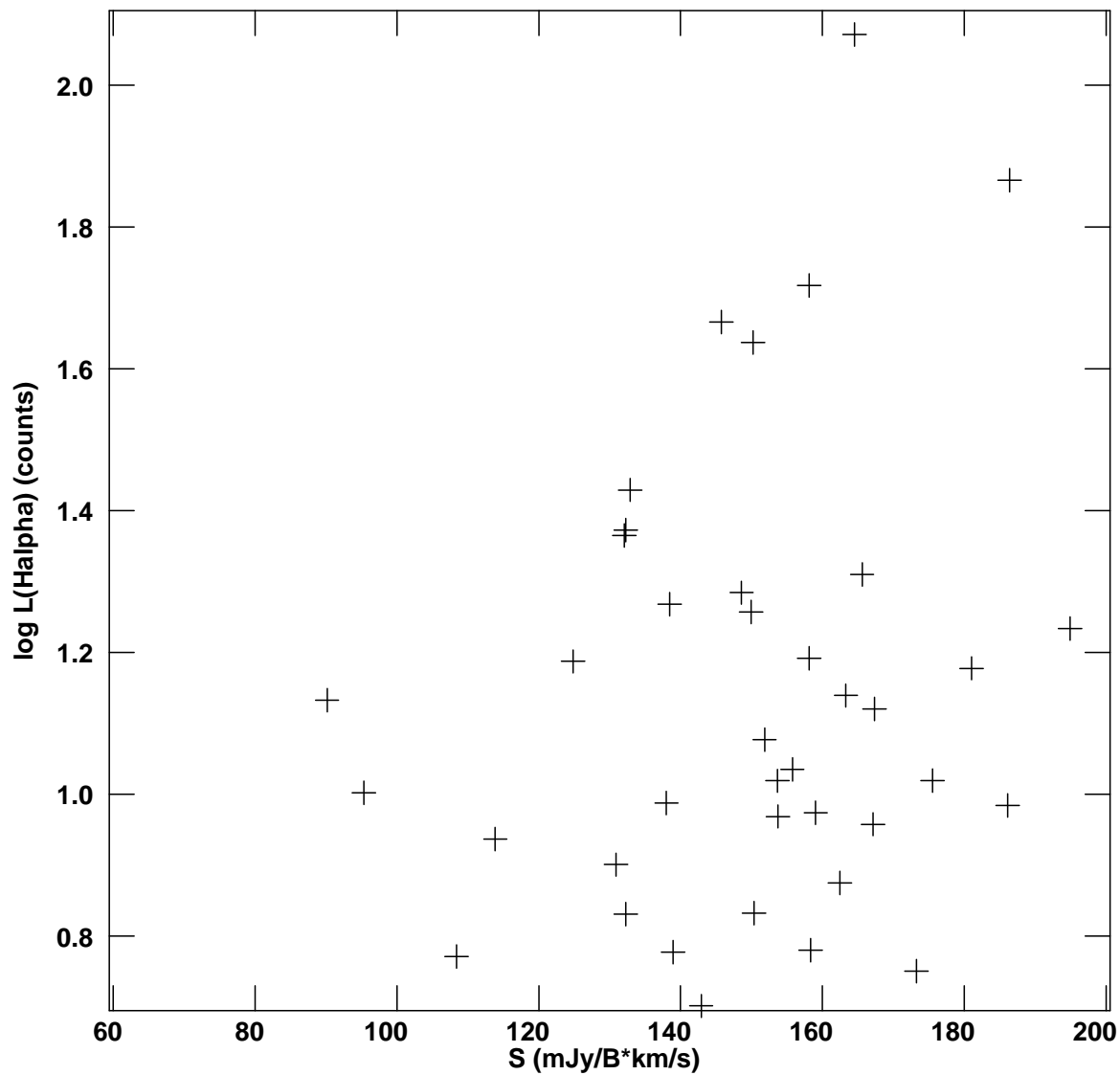}
\caption{Brightnesses of individual pixels are compared in \protect\HI\
  and \protect\ha.  The integrated C-configuration \protect\HI\ map was
  geometrically matched to the \protect\ha\ image, and then they were
  both averaged 20$\times$20 to produce a pixel scale of
  9.8\protect\arcsec, close to the beam of the original C-configuration
  HI map (14\protect\arcsec$\times$11\protect\arcsec).  The x-axis is
  the \protect\HI\ pixel brightness in mJy Beam$^{-1}$ \protect\kms\
  where 60 mJy Beam$^{-1}$ \protect\kms\ corresponds to a column density
  of $4.3\times10^{20}$ \protect\coldens.  The y-axis is the logarithm
  of the \protect\ha\ pixel brightness in counts in the image where 1
  count corresponds to an \protect\ha\ luminosity of $1.95\times10^{33}$
  ergs s$^{-1}$, corrected for reddening.
\label{fig:imvim}}
\end{figure}

In Figure \ref{fig:imvim} we compare the integrated \HI\ and \ha\
brightnesses of individual pixels in the galaxy. The integrated C-configuration
\HI\ map was geometrically matched to the \ha\ image. Then both were
averaged 20$\times$20 in order to approximately match the \HI\ beam of
the C-configuration map.  The beam size is 14\arcsec$\times$11\arcsec; and the
pixel scale of the averaged maps was 9.8\arcsec.  The x-axis of this
figure goes from an observed \HI\ column density of 4.3$\times10^{20}$
\coldens\ to 1.4$\times10^{21}$ \coldens.  The y-axis goes from an \ha\
luminosity of 9.8$\times10^{33}$ ergs s$^{-1}$ to 2.5$\times10^{35}$
ergs s$^{-1}$, corrected for reddening (1 count in the \ha\ image is
1.95$\times10^{33}$ ergs s$^{-1}$; we have not converted this to a
surface brightness by dividing by the area of the pixel).  \ha\ is
corrected for reddening assuming
E(B$-$V)$_t$$=$E(B$-$V)$_f$+0.1$=$0.155.  With a Cardelli \et\ (1989)
reddening curve, A$_{H\alpha}$$=$0.48.  There is a slight trend that the
highest \ha\ luminosities are found at the highest observed \HI\ column
densities. This is similar to what was found for NGC 2366 (Hunter \et\
2001).

\placefigure{fig:imvim}

However, in DDO 43 there are also \HII\ regions found at lower column
densities and not associated with any apparent peak in \HI.  There are
two small \HII\ regions in the northeast part of the galaxy that lie at
\HI\ column densities of about $7\times10^{20}$ \coldens.  In addition
there are several \HII\ regions lying in the central \HI\ holes, also at
column densities of $7\times10^{20}$ \coldens.  One would expect that
cloud formation in these regions would be harder because of the lower
column densities, especially so for the outer \HII\ regions where the
critical gas density for gravitational instabilities is dropping
(Safronov 1960, Toomre 1964).

\subsection{The Holes in the Gas}

We have identified four holes in the \HI\ in DDO 43. The largest has a
diameter of 850 pc and the smallest is only 110 pc in size. Holes have
been found in the \HI\ maps of other disk galaxies as well. Im galaxies
with cataloged gas holes include IC10 with 8 holes (Wilcots \& Miller
1998), DDO 47 with 19 holes (Walter \& Brinks 2001), DDO 50 with 51
holes (Puche \et\ 1992), DDO 81 with 48 holes (Walter \& Brinks 1999),
NGC 6822 with one hole (de Blok \& Walter 2000), and the LMC with many
holes (Kim \et\ 1999).  Some of the holes in these galaxies reach 2.0
kpc in diameter, but most are several hundred parsecs in size. Thus, the
\HI\ holes in DDO 43 are typical in size. However, we have not detected
the signature of expansion in any of DDO 43's holes.  Therefore, they
must be relatively old.

DDO 43's largest hole (``Hole 1'') resembles that surrounding the OB
association NGC 206 ($=$OB78) in M31 (Brinks 1981). The \HI\ hole around
NGC 206 has clearly been produced by the mechanical energy input from
the concentration of massive stars in the enclosed OB association.  Not
only do massive stars explode as supernovae, but they dump a comparable
amount of energy into their surroundings over their lifetimes in the
form of strong winds. Brinks estimates that 2$\times 10^6$ \msun\ of
\HI\ is ``missing'' from the NGC 206 hole. If we assume that the column
density prior to the formation of DDO 43's hole was the same as the
average we see in the knots (1.3$\times 10^{21}$ \acm), and covered the
area out to the current 0.8$\times 10^{21}$ \acm\ contour, then the
amount of \HI\ that is now ``missing'' is also about 2$\times 10^6$
\msun.  The similarity between Hole 1 and the hole around NGC 206
implies that a large OB association like NGC 206 is capable of producing
Hole 1. While NGC 206 is a substantial OB association and would have
dominated DDO 43's optical appearance when it was young, it is not an
unusual or extreme event and would not have contributed more than 0.08\%
to the mass of stars in the galaxy.  The smaller holes in DDO 43 would
have been produced by more modest OB associations.


We measured the UBV colors of the stellar population within Hole 1 using
a 5.7\arcsec\ radius aperture centered on the \HI\ position.  The colors
are (B$-$V)$_0=0.35\pm0.04$ and (U$-$B)$_0=-0.20\pm0.05$. (B$-$V)$_0$ is
nearly the same as that of the galaxy as a whole, but (U$-$B)$_0$ is 0.1
mag redder.  These colors do not yield a consistent age for the stellar
population, assuming a single star-forming event, perhaps because of
contamination by a background galaxy population. However, the slightly
redder color of the hole compared to the rest of the galaxy is
consistent with an average age of the stellar population in the hole
that is a little older than the surroundings. The star-forming event
that formed Hole 1 must be old enough that we do not, at least at our
modest spatial resolution, see an obvious remnant of the OB
association. On the other hand, the age must be young enough that the
\HI\ hole is still distinct and has not yet filled back in.

Wada, Spaans, \& Kim (2000) have suggested a mechanism for producing
holes in the interstellar medium of Im galaxies that does not depend on
the energy input from massive stars. They argue that holes can be caused
by nonlinear evolution of the multi-phase interstellar medium.  This
search for a non-stellar production mechanism for holes was motivated by
the lack of evidence for aging OB associations in some of the holes in
DDO 50.  The hole in NGC 6822 also presents a problem because it is
located in the outer parts of the optical galaxy.  While we cannot rule
out this sort of mechanism for the production of the holes in DDO 43,
the modest star-forming events needed to produce these holes are likely
events in the evolution of an Im galaxy like DDO 43. Therefore, we do
not need to invoke other mechanisms.

The gas holes in DDO 43 are located near the center of the galaxy.  A
ridge of \HI\ surrounds the holes to the north on three sides and a
large \HI\ complex sits to the south. The ridge and complex together
give one the impression of a ring that is 2.7 kpc $\times$ 2.0 kpc.
Because the center of the ring is composed of several distinct holes
rather than a single one, it seems likely that the ring is a consequence
of multiple star-forming events that have taken place near the center of
the galaxy.  The gas shells that resulted have run into each other to
give the appearance we have today. Furthermore, this would be a natural
means for producing the enhanced \HI\ density that we see in the
ring. The ring is where most of the star formation is taking place
today. In this scenario, the star formation in the ring would represent
a second generation and would be an example of star-induced star
formation (Gerola, Seiden, \& Schulman 1980).

DDO 43 is not alone in containing a large gas ring. DDO 88 contains a
beautifully complete ring 3.0 kpc in diameter (Simpson \et\ 2005).  In
addition NGC 2366 appears to have a partially broken ring that is 5.5
kpc in diameter (Hunter \et\ 2001), and M81dwA and SagDIG have partial
rings of 2 kpc and 1 kpc diameter (Sargent, Sancisi, \& Lo 1983; Young
\& Lo 1997).

\section{Summary}

DDO 43 optically resembles other dwarf galaxies that fall at the faint
end of the Im distribution. It is relatively isolated, with 270
pc to the nearest Im and 1.6 Mpc to the nearest large galaxy (a
spiral). It probably hosts a large stellar bar: it
exhibits boxy elliptical isophotes with a small shift in position angle
proceeding from the inner to outer regions, and there is an almost
90\arcdeg\ misalignment between the optical (bar) axis and the kinematic
axis determined from the \HI\ data. Although the bar is large at 0.8R$_{25}$,
galaxy-sized bars are often seen in Ims that have bars. 
The UBV surface photometry reveals an exponential disk with a normal
central surface brightness and an not-unusual scale length of 430
pc. The galaxy's integrated colors are normal, and 
like many small systems it exhibits little color gradient across the
disk. There are six star clusters visible on the V image. Based on their
colors, evolutionary track models indicate that only one is possibly
large enough to be considered a ``populous'' cluster; all the others are
relatively small.

In the \HI, DDO 43's  extended \HI\ is
emphasized by its large R$_{HI}/$R$_{H}$ ratio. There is an extensive,
low-level \HI\ envelope containing a high density ridge surrounding the
center of the galaxy. This ridge contains higher density knots and
lower density depressions in the \HI. 44\% of the total \HI\ mass is
located within this ridge, defined as the region having a column density
in excess of 1$\times 10^{21}$ \acm. We have identified six knots with
N$_{HI} \geq 1.2\times 10^{21}$ \acm; the largest knot is roughly
750$\times$400 pc across. There are four ``holes,'' the largest, located
just north of the largest \HI\ knot, is 850 pc across. Holes of this
size are often detected in Ims and are thought to be the result of star
formation activity.

The largest hole in DDO 43 resembles the one formed by star formation
that surrounds the OB association NGC 206 in M31. However, there are no
young star associations in DDO 43's holes; the stellar colors are
slightly redder than those in the rest of the galaxy and they do not
appear to be expanding, so they must be relatively old. The knots in the
high density ridge form a broken ring interspersed with the holes. The
structure of knots and holes we see today is most likely the result of
the overlapping of expanding \HI\ shells caused by the stellar 
winds and supernovae explosions from earlier star formation episodes.
Most of the current star formation is taking place in the ridge, possibly
induced by the sweeping up of gas as the holes were formed. 

DDO 43's current star formation rate is normal, falling in the middle range
for Ims. There are several moderate-size \HII\ regions scattered around
the galaxy, but there is on average less star formation activity in the
inner 0.75 kpc than expected based on a comparison of the
azimuthally-averaged \ha\ and V-band surface densities. Most of the
\HII\ regions are located in areas of high \HI\ surface density, and
many of those are located near local peaks in the \HI, as is seen in
other Ims. A few of the \HII\ regions are associated with lower \HI\
surface density areas however, including the \HI\ holes.

Kinematically, there is nothing strikingly unusual about DDO 43. It is
undergoing the primarily solid-body rotation seen in many dwarfs, with
an S-distortion visible in the velocity field, possibly indicating the
presence of a warp in the gas disk. The rotation curve levels off and
possibly turns over. The maximum rotation velocity is 25 \kms, which is
higher than the original estimate. It was the low rotation velocity for
its luminosity that caused this galaxy to be classified as a candidate
``fossil'' tidal dwarf. Dwarf galaxies formed from tidal interactions
are predicted to contain less dark matter than primordially formed
dwarfs, so should have lower total masses and subsequently, lower
rotation speeds. However, the revised velocity estimate puts it back in
the normal area of the Tully-Fisher relation for Ims, so there is no
reason to think that DDO 43 formed from an interaction.

The dispersion velocities in the \HI\ indicate an average of $\sim$ 10
\kms, which is normal. There is one area of higher dispersion; this is
associated with the large hole. The \HI\ spectra through the hole
indicate that the gas motions are complex, with most of the spectra
exhibiting four peak components. If the hole was produced by star
formation activity, it would seem that the gas is still somewhat
unsettled. 

Overall, DDO 43 is a fairly typical gas-rich, optically faint, small
dwarf galaxy. It has an extensive gas disk and a high M$_{HI}$/L$_B$
ratio. The \HI\ has been sculpted into a high density ridge forming a
partial ring that contains knots and holes. This is most likely the
result of multiple episodes of star formation at various times. The
mechanical energy deposited in the gas has carved out regions of lower
gas density as gas was swept up into expanding shells. High density
regions occur where the shells have overlapped. As a result, star
formation was triggered in these areas as current star formation
activity is located there. The shells are not currently expanding, so
the events that formed them must have happened long ago. The current
star formation rate is modest; coupled with the large gas reservoir and
modest sizes of the older clusters seen in V, it seems that DDO 43 has
evolved slowly, and will continue to do so for the next several Gyr.

\acknowledgments

We wish to thank Emily Bowsher for examining the color ratio images of
DDO 43 as part of the 2003 Research Experiences for Undergraduates
program of Northern Arizona University.  DAH obtained partial support
for this research from the Lowell Research Fund.  Additional support to
DAH, support for TEN, and support for travel to Flagstaff for CES came
from grant AST-0204922 from the National Science Foundation.

This research has made use of the NASA/IPAC Extragalactic Database (NED)
which is operated by the Jet Propulsion Laboratory, California Institute
of Technology, under contract with the National Aeronautics and Space
Administration.

\clearpage

\begin{deluxetable}{lcccc}
\tablecaption{VLA Observations
\label{tab:HI}}
\tabletypesize{\small}
\tablewidth{0pt}
\tablehead{
\colhead{} 		& \colhead{D configuration}	
				& \colhead{C configuration}	
						& \colhead{C$+$D$_{UN}$\tablenotemark{a}}  
								& \colhead{C$+$D$_{NA}$\tablenotemark{a}}
}
\startdata

Observation Date	& 1996 Sept.\ 20& 2000 April 9	& \nodata	& \nodata\\	
Time on Source~(min)	  & 274		& 268 		& \nodata	& \nodata\\
Bandwidth~(MHz)		  & 1.56 	& 1.56 		& 1.56		& 1.56\\
No.\ of Channels	  & 128		& 128		& 128		& 128\\				    
Velocity Resolution (\kms)& 2.6 	& 2.6 		& 2.6 	    	& 2.6\\			    
Beam Size\tablenotemark{b} \ (\arcsec) 
			  & $54.8 \times 52.7$
					& $14.0 \times 11.0$
							&$20.3 \times 14.4$
                                                                        &$24.5 \times 17.1$\\
Single channel r.m.s.~(mJy B$^{-1}$)& 0.81	& 1.05 		& 1.07 		& 1.05 \\
Brightness Temperature~(K)& 0.24	& 3.15		&\nodata	&\nodata \\
\enddata
\tablenotetext{a}{Combined data set from both configurations}
\tablenotetext{b}{Natural weighting for D and C$+$D$_{NA}$; uniform weighting with
robust factor of +1 for C and C$+$D$_{UN}$}
\end{deluxetable}

\clearpage

\renewcommand{\arraystretch}{.6} 
\begin{deluxetable}{lc}
\tablecaption{Summary of Integrated Properties.
\label{tab:int}}
\tabletypesize{\small}
\tablewidth{0pt} 
\tablehead{ 
\colhead{Parameter} & \colhead{Value} 
}
\startdata 
   D (Mpc) \dotfill 				& 5.5 \\ 
   M$_{HI}$ (M\protect\solar) \dotfill 		& 9.2$\times10^7$ \\
   E(B$-$V)$_f$\tablenotemark{a} \ \dotfill 	& 0.055 \\
   R$_{25}$ (arcsec, kpc) \dotfill 			& 37, 0.99 \\
   R$_H$ (arcsec, kpc) \dotfill 			& 53, 1.4 \\
   R$_{HI}$\tablenotemark{b} \ (arcsec, kpc) \dotfill	& 146, 3.9 \\
   $\mu_0$ (V-band, mag arcsec$^{-2}$) \dotfill & 22.4$\pm$0.2 \\
   R$_D$ (pc) \dotfill 				& 430$\pm$50 \\
   M$_{V_0}$ (R$=$57\protect\arcsec) \dotfill 	& $-14.31\pm0.02$ \\
   (U$-$B)$_0$ (R$=$57\protect\arcsec) \dotfill	& $-0.31\pm0.04$ \\
   (B$-$V)$_0$ (R$=$57\protect\arcsec) \dotfill	& $0.31\pm0.02$ \\
   log L$_{H\alpha,0}$ (ergs s$^{-1}$) \dotfill	& $38.80\pm0.006$ \\
   SFR\tablenotemark{c}~
      (M\protect\solar/yr$^{-1}$) \dotfill 	& 0.0037 \\
   log SFR/area\tablenotemark{c}~ 
      (M\protect\solar\ yr$^{-1}$ kpc$^{-2}$) \dotfill 
                                                & $-$2.91 \\
   M$_{HI}$/L$_B$ 
      (M\protect\solar/L\protect\solar) \dotfill& 1.64 \\

\enddata 
\tablenotetext{a}{E(B$-$V)$_f$ is foreground reddening due to
the Milky Way (Burstein \& Heiles 1984).  For the stars in DDO 43, we
assume an additional internal reddening of 0.05 magnitude; for the
\protect\HII\ regions we assume an additional internal reddening of 0.1 
magnitude, consistent with Balmer decrement observations of emission
nebulae in DDO 43 (Hunter \& Hoffman 1999).}  
\tablenotetext{b} {Measured to N$_{HI} = 5\times 10^{19}$ \acm;
  corrected for the beam-size.}
\tablenotetext{c}{Star
formation rate derived from L$_{H\alpha}$ using the formula of Hunter \&
Elmegreen (2004) that integrates from 0.1 M\protect\solar\ to
100 M\protect\solar\ with a Salpeter (1955) stellar initial mass
function.  The area is $\pi$R$_{25}^2$.}
\end{deluxetable}

\clearpage

\renewcommand{\arraystretch}{1} 
\begin{deluxetable}{ccccccc}
\tablecaption{Star cluster properties.
\label{tab:clus}}
\tablewidth{0pt} 
\tablehead{ 
\colhead{Cluster} & \colhead{M$_{V_0}$} &
\colhead{U$-$B$_0$} & \colhead{B$-$V$_0$} & \colhead{Age} &
\colhead{M$_V$} &
\colhead{Nebula?\tablenotemark{b}}\\
\colhead{} &\colhead{} & \colhead{} & \colhead{} &\colhead{(Myr)} &
\colhead{(10 Myr)\tablenotemark{a}} & \colhead{} 
} 
\startdata 1 & $-8.63\pm0.15$ &
$0.25\pm0.21$ & $-0.55\pm0.21$ & 7--70 & $-7.4$ to $-10.2$ & N \\
 2 &
$-6.89\pm0.24$ & $0.03\pm0.32$ & $-0.89\pm0.24$ & 3--20 & $-6.0$ to
$-7.6$ & N \\
 3 & $-8.66\pm0.05$ & $0.18\pm0.07$ & $-0.43\pm0.07$ &
7,15 & $-7.5$, $-8.8$ & Y \\
 4 & $-9.28\pm0.07$ & $0.56\pm0.12$ &
$-0.66\pm0.13$ & 8--15: & $-8.6$ to $-9.4$ & Y \\
 5 & $-8.07\pm0.13$ &
$0.05\pm0.18$ & $-0.68\pm0.15$ & 5--30 & $-7.1$ to $-9.1$ & Y \\
 6 &
$-8.50\pm0.09$ & $0.11\pm0.11$ & $-0.98\pm0.08$ & 5: & $-7.5$ & Y \\

\enddata 
\tablenotetext{a}{The M$_V$ that the cluster will or would have
had at an age of 10 Myrs, determined from the ages and the cluster
evolutionary track of Leitherer et al.\ 1999 for a metallicity of
$Z=0.008$.}  \tablenotetext{b}{The presence of nebular emission suggests
that the younger age in a range is more appropriate.}
\end{deluxetable}


\clearpage

\clearpage

\clearpage

\clearpage

\clearpage

\clearpage

\clearpage

\clearpage

\clearpage

\clearpage

\clearpage

\clearpage

\clearpage

\clearpage

\clearpage

\clearpage

\clearpage

\clearpage

\clearpage

\clearpage

\clearpage

\clearpage

\end{document}